# Single-walled $CrI_3$ nanotubes encapsulated within multiwall Carbon Nanotubes

[1]*Ihsan Çaha*, [1]*Aqrab ul Ahmad*, [1]*Loukya Boddapatti*, [1]*Manuel Banõbre*, [1]*António T. Costa*, [2]*Andrey N. Enyashin*, [3]*Weibin Li*, [3]*Pierluigi Gargiani*, [3]*Manuel Valvidares*, [1]*Joaquín Fernández-Rossier**, [1]*Francis Leonard Deepak**

[1]Ihsan Çaha, Aqrab ul Ahmad, Loukya Boddapatti, Manuel Banõbre, António T. Costa, Joaquín Fernández-Rossier, Francis Leonard Deepak
INL - International Iberian Nanotechnology Laboratory,
Avenida Mestre José Veiga s/n,
Braga 4715-330,
Portugal
E-mail: joaquin.fernandez-rossier@inl.int, Leonard.Francis@inl.int

[2]Andrey N. Enyashin
Institute of Solid State Chemistry,
UB RAS,
620990 Ekaterinburg,
Russian Federation

[3]Weibin Li, Pierluigi Gargiani, Manuel Valvidares
ALBA Synchrotron Light Source,
E-08290 Cerdanyola del Vallès, Barcelona,
Spain

Keywords: 1D van der Waals heterostructures, $CrI_3$ nanotubes, carbon nanotube encapsulation, curvature-induced magnetism

$CrI_3$ is a layered ferromagnetic insulator that has recently attracted enormous interest as it was the first example of a stand-alone monolayer ferromagnet, paving the way towards the study of two-dimensional magnetic materials and their use as building blocks of hybrid van der Waals layered heterostructures. Here we go one step down in the dimensionality ladder and report the synthesis and characterization of a tubular one-dimensional van der Waals heterostructure where $CrI_3$ nanotubes are encapsulated within multiwall carbon nanotubes (MWCNT), integrating a magnetic insulator and a conductor. By means of the capillary filling of MWCNT, we obtained single-walled $CrI_3$ nanotubes with diameters ranging between 2 nm and 10 nm, with an average of 5.3 nm. Through aberration-corrected transmission electron microscopy, X-ray magnetic circular dichroism (XMCD) spectroscopy, and first-principles calculations, we explored the fundamental physics and magnetism of these 1D van der Waals heterostructures. These findings pave the way towards the exploration of non-collinear magnetic states in tubular geometries, driven by the interplay of magnetic anisotropy and curvature.

## 1. Introduction

2D van der Waals structures (vdWs) can be designed by combining two or more atomically thin layers stacked on top of each other with no chemical binding other than van der Waals forces, analogously to how different pieces are stacked in a LEGO game.[1] As a consequence, the combination of layers offers a wide range of possibilities to create a vast amount of new hybrid materials and devices with very unusual and interesting properties with the potential to significantly impact different fields like optoelectronics,[2] various quantum technologies,[3] valleytronics,[4] spintronics,[5–7] etc. The potential for this strategy is enormous, given the discovery of non-trivial emergent properties that depend strongly on the twist angle between the layers.[8] In contrast to the 2D van der Waals stacks, whose experimental exploration is more than 10 years old by now[1], the formation of 1D vdWs heterostructures (1D vdW Hs), a class of materials where different atomic layers are coaxially stacked, is an intriguing and new concept.[9] The interplay between magnetism and curvature is a very active field of research in artificial structures at the micron scale.[10,11]

Magnetic 2D crystals were discovered relatively recently, with the observation of antiferromagnetic order in $FePS_3$ monolayers[12] and ferromagnetic order in monolayer $CrI_3$.[13] $CrI_3$ monolayers have elicited a gigantic interest for several reasons.[14] First, their strong off-plane magnetic anisotropy stabilizes magnetic order in two dimensions.[15] Second, they can be integrated into multilayer van-der-Waals heterostructures such as spin-filter magnetic tunnel junctions, making it possible to use tunneling electrons to probe their properties, or next to transition metal dichalcogenides, leading to spin-proximity effects. The electrically controllable spin proximity effect has also been predicted to occur between graphene and graphene bilayers and $CrI_3$, opening the way to new device concepts.[16] It is thus apparent that $CrI_3$ monolayers are fascinating both in their own right and because of their interactions with other layers in van der Waals heterostructures. In parallel, the groundbreaking research on 2D $CrI_3$ has spurred the discovery of new 2D magnetic materials, obtained from layered bulk crystals,[17,18] and the fabrication of planar van der Waals heterostructures that combine these materials with semiconducting and superconducting materials. This has led to the observation of new physical phenomena, including claims of topological superconductivity.[19] Additionally, recent experimental studies on curvilinear magnetism, such as magnetic nanoshells[20,21], curved graphene[22], and freestanding layers of antiferromagnetic oxides[23] have further demonstrated the significant role of curvature in influencing magnetic properties.

$CrI_3$ monolayer magnetic nanotubes with a nanometer radius have been predicted to transition to a radial magnetization state[24] on account of the interplay between the perpendicular magnetic anisotropy of the layered material and curvature. Despite efforts on the synthesis of nanotubes for more than a decade now, compared to their two-dimensional "flat" analogues, the amount of single-walled (inorganic) nanotubes (SWNTs) reported to date is limited (C, BN are the exceptions) because their multiwalled counterparts are favored during synthesis, thus posing a great challenge and hurdle for investigating novel phenomena, properties, and applications. The formation of inorganic SWNTs is associated only with their high strain energies, not naturally involving any interlayer interaction. Hence, narrow SWNTs of inorganic compounds have been predicted to be less stable than their multiwalled counterparts.[25] 1D van der Waals heterostructures have been demonstrated in $PbI_2$ within the cavities of carbon nanotubes[26,27] and have been expanded to other inorganic systems employing the CNT template approach.[28] Other 1D vdW heterostructures that have been successfully demonstrated and whose structure elucidated include those of $BiI_3$ and $GdI_3$.[29–31]

Here, we report the successful realization of 1D monolayer $CrI_3$ nanotubes, confirmed by advanced electron microscopic characterization and element-specific synchrotron spectroscopy, supplemented with calculations to uncover the nature of the nanoscale magnetic state. A chemical vapor transport technique was utilized for the growth of the one-dimensional van der Waals heterostructures of monolayer $CrI_3$ nanotubes encapsulated within multiwalled Carbon nanotubes (MWCNTs). Morphological and chemical analyses were conducted to confirm the formation of a single wall of $CrI_3$ nanotube without any impurities inside the MWCNTs. Furthermore, the magnetic properties of these $CrI_3$ 1D van der Waals heterostructures (vdWHs) were investigated by using XMCD and SQUID techniques. Density functional theory (DFT) calculations were carried out to complement the experimental results regarding the morphology of the tubular $CrI_3$ structure. In addition, both previous DFT calculations[24] as well as a model Hamiltonian presented here, predict the emergence of a radial magnetization state in our single walled $CrI_3$ nanotubes.

## 2. Results and Discussion

The $CrI_3$ tubes encapsulated in MWCNT were obtained by the capillary filling method (see **Figure 1a** and also the methods section). The nanotubes can be observed using high-angle annular dark-field scanning transmission electron microscopy (HAADF-STEM), high-resolution TEM (HRTEM) and annular bright field STEM (see **Figures 1, 2, 3**, and **S2**), demonstrating the encapsulation of $CrI_3$. The detailed morphology and structural characterization of individual encapsulated $CrI_3$ nanotubes were achieved by using a double-corrected FEI Titan G3 Cubed Themis operated at 200 kV. In **Figures 1b** and **1c**, the corresponding HAADF-STEM images clearly reveal the formation of the single-walled $CrI_3$ nanotubes encapsulated within the host MWCNTs, whereas **Figure 1d** shows the high-resolution TEM image of the single-walled $CrI_3$ nanotube encapsulated within a very clear multiwalled CNT. In the HAADF-STEM images, the brighter regions correspond to $CrI_3$ within the darker CNTs region due to the atomic number (Z) dependence of the image contrast. Since the contrast of an atomic column in the incoherent HAADF-STEM image is proportional to Z, this technique enables us to directly distinguish the heavier $CrI_3$ atoms from the lighter carbon atoms of the outer layers. The formed single-walled $CrI_3$ nanotubes are single crystalline in nature, equivalent to a 2D layer of $CrI_3$ wrapped into a cylindrical form, resulting in a strain gradient in the curvature of the nanotubes as shown in **Figure S3**. However, these strains are smoothly distributed, as seen in both **Figure S3a** and **S3b**, without the sharp discontinuities typically associated with dislocations.

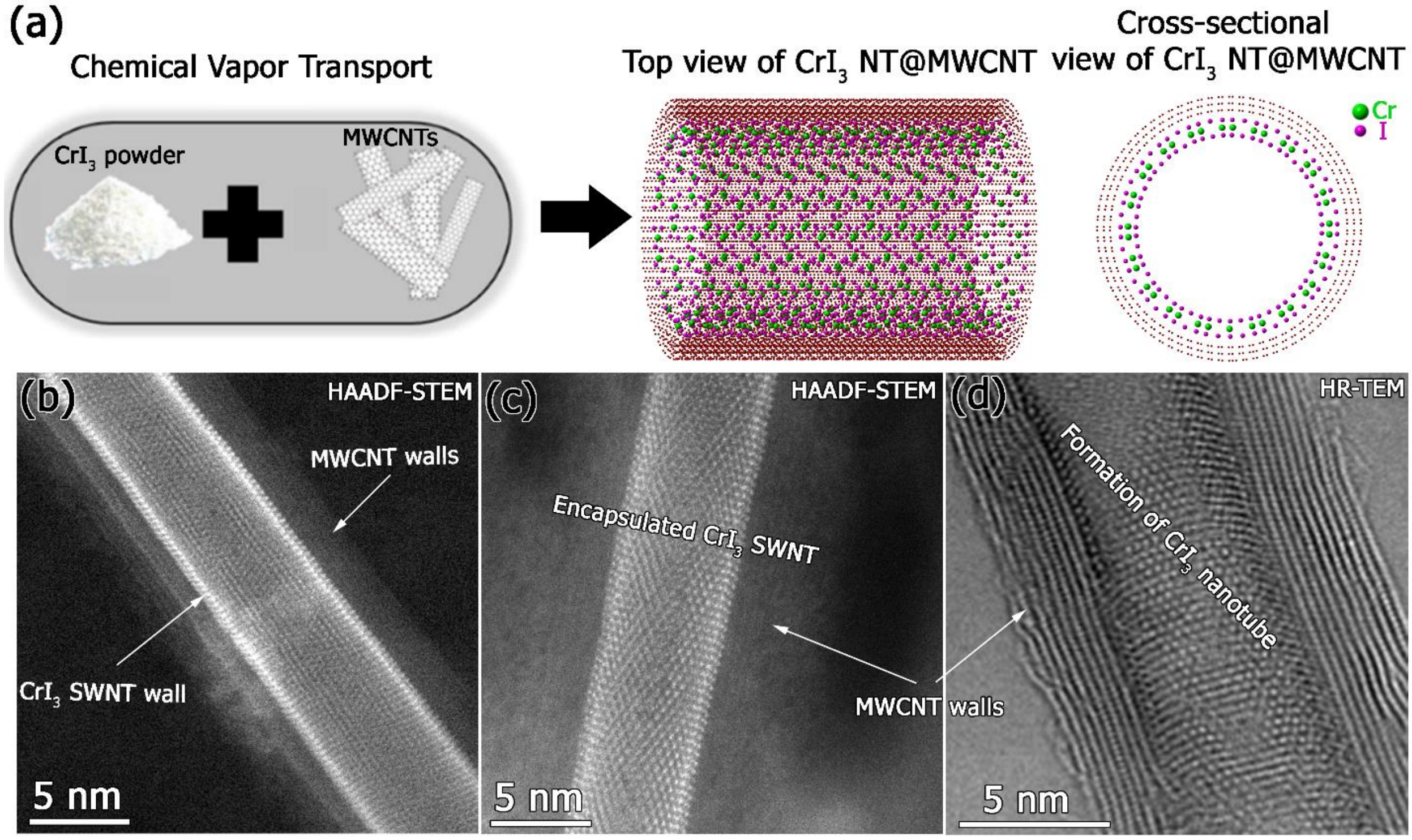

**Figure 1**. (a) Scheme of the preparation method, showing the filling of CNTs with $CrI_3$ phases, together with top-view and cross-sectional view of CNTs filled with $CrI_3$ single-walled nanotubes (violet = $CrI_3$ single-walled NTs and red = CNTs). High-resolution HAADF-STEM images in (b), (c), and HR-TEM image in (d) clearly show the formation of the single-walled $CrI_3$ nanotubes encapsulated within the host CNTs.

To verify the chemical composition of the encapsulated structures, we performed energy-dispersive X-ray spectroscopy (EDS) mapping, as shown in **Figure 2**. The elemental maps confirm that the filled nanotubes consist of Cr and I, indicating the successful formation of $CrI_3$ within the carbon nanotubes. Quantitative analysis of the EDS spectrum (**Figure 2b**) reveals an elemental ratio of approximately 24 at.% Cr and 76 at.% I, which closely matches the expected stoichiometry of $CrI_3$ (1:3 ratio). Furthermore, the EDS line scan (**Figure 2c**), taken along the red line marked in the HAADF-STEM image (**Figure 2a**), displays distinct elemental distributions consistent with a hollow tubular structure. The signal intensity profiles for Cr and I align well with the spatial location of the $CrI_3$ wall, confirming the formation of a single-walled $CrI_3$ nanotube encapsulated within the MWCNT.

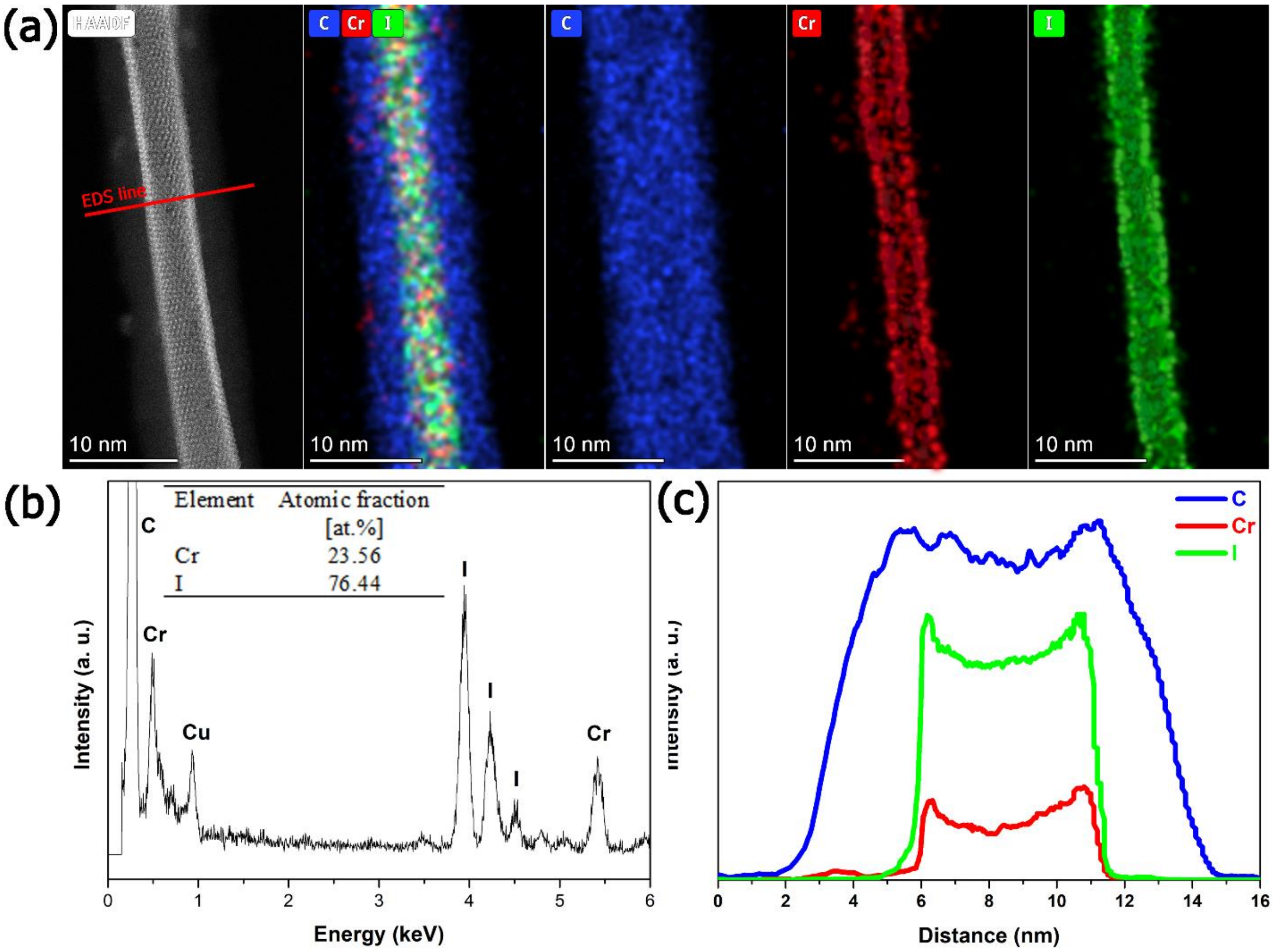

**Figure 2**. EDS analysis showing the chemical constituents of the filled single-walled $CrI_3$ nanotube: (a) HAADF-STEM image, and EDS map of carbon, chromium, and iodine together and individual, (b) EDS spectrum taken from the HAADF image, and (c) EDS line scan taken from the region shown in (a) reflecting the tubular shape of the structures, along with the atomic percentage fraction obtained (inset in b), for the $CrI_3$ nanotube compositional analysis.

The high-resolution HAADF-STEM analyses together with statistical diameters of MWCNTs and single-wall $CrI_3$ nanotubes are presented in **Figure 3**. A high-resolution HAADF-STEM image of an encapsulated single walled $CrI_3$ nanotube (**Figure 3b**) along with its Fast Fourier Transform (FFT) image (**Figure 3d**) reveals the zigzag configuration of $CrI_3$ nanotubes. **Figures 3e** and **3f** show the diameter distributions of MWCNTs and encapsulated $CrI_3$ nanotubes, respectively. The average outer diameter of the MWCNT was 9.3 nm, with a diameter distribution range between 4 - 22 nm, while its average inner diameter was 5.6 nm with a diameter distribution range between 2 - 10 nm. Although the average diameter of the $CrI_3$ nanotubes (5.3 nm) was smaller than the average diameter of MWCNT, its distribution was in the range of the inner diameters of CNT. This distribution range and the preferential formation of nanotubes at larger diameters are in good agreement with reports for other metal halides[26,29,30]. In addition, optical Raman microscopy was used at two different wavelengths to investigate charge transfer in the vdW heterostructures. The Raman spectra of pristine MWCNTs and CNT-filled $CrI_3$ samples (**Figure S4** and **Table 1**) exhibited no significant shifts or shape changes observed in the G-band (1588 $cm^{-1}$) and 2D mode, suggesting an absence of detectable charge transfer in the global Raman response.[32,33] However, it is crucial to recognize that charge transfer in these systems is highly localized, predominantly occurring between the first carbon wall of the MWCNT and the encapsulated single-walled $CrI_3$ nanotube. Our Density functional theory (DFT) calculations (**Figure S13** and **Table S2**) predicted a small charge transfer from the $CrI_3$ nanotube to the innermost CNT wall, treated as a graphene-like structure. To experimentally validate this prediction, we performed 4D-STEM charge density mapping, which directly confirmed the presence of localized charge transfer between the $CrI_3$ nanotube and the first CNT wall, as given in **Figure S5**. Interestingly, our analysis further revealed that the subsequent CNT layers act as an effective barrier, preventing charge redistribution beyond the first few layers. As a result, by the time the signal traverses 5 to 10 CNT layers, no discernible charge transfer remains. This confinement effect explains the lack of observable Raman shifts, as Raman spectroscopy, being a global probe, primarily captures the cumulative response of all CNT layers rather than the localized interfacial interactions. Our findings highlight the importance of employing local charge mapping techniques such as 4D-STEM to accurately assess charge transfer phenomena in multi-walled heterostructures.

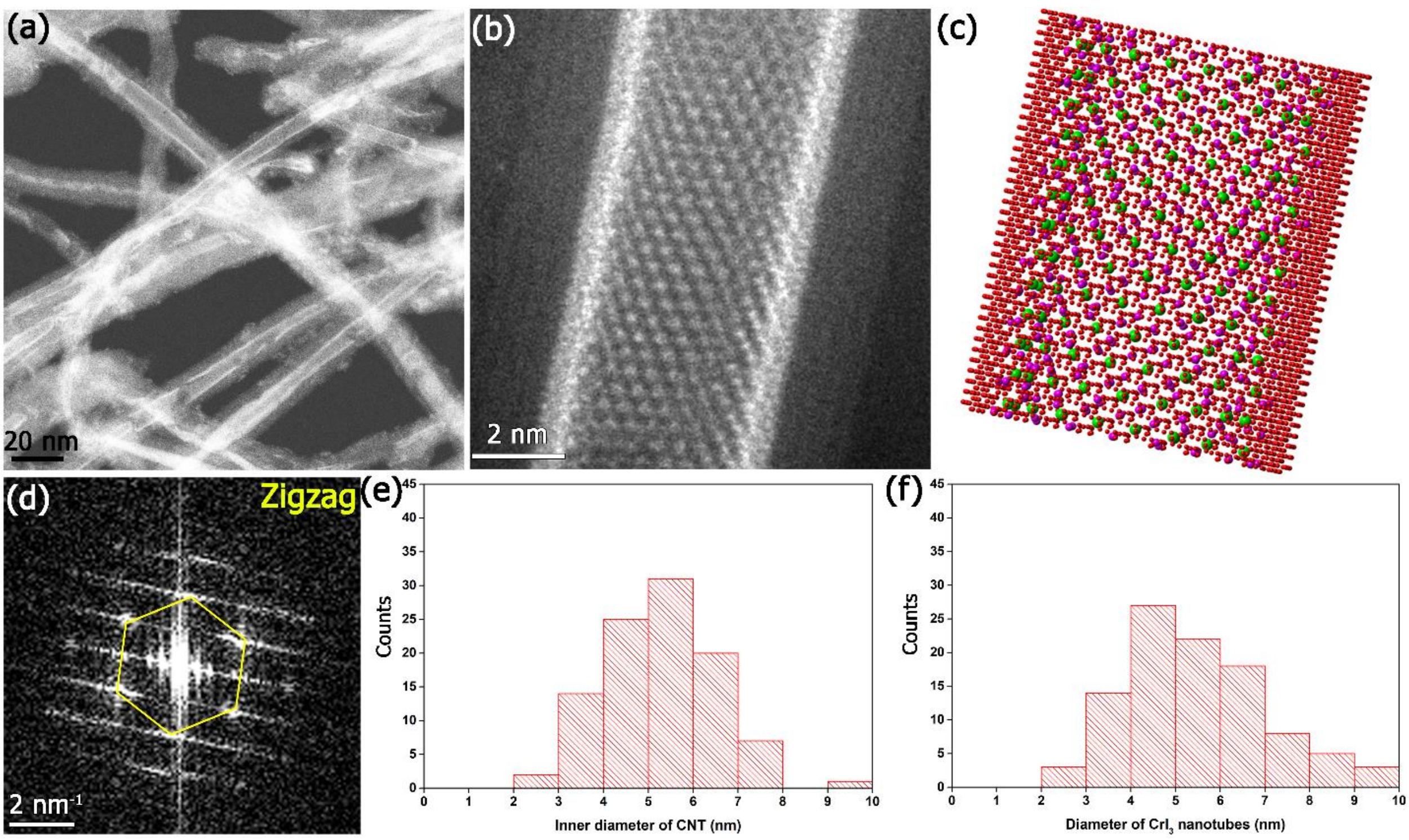

**Figure 3**. (a) Overview of the filled CNTs in a HAADF-STEM image, (b) High magnification HAADF-STEM image of a $CrI_3$ SWNT@MWCNT, and (c) The atomic structure model of the $CrI_3$ nanotube in the panel (b), and (d) FFT taken from image (b) showing the zigzag nanotube formation, (e) Statistical distribution of inner MWCNTs and (f) diameter distribution of the single walled $CrI_3$ nanotubes.

The magnetic behavior of both the $CrI_3$ powder precursor and the $CrI_3$ 1D-heterostructures was investigated by means of Superconducting Quantum Interference Device (SQUID) magnetometry, complemented with element-specific X-ray Absorption Spectroscopy (XAS) and X-ray magnetic circular dichroism.[34–36] Our hysteresis data for the powder precursor show a ferromagnetic M(T) curve (**Figure S6**) with a magnetic moment per Cr atom of 0.6 Bohr magneton (BM), 5 times lower than the values reported for single crystal samples.[37,38] (which can be due to the applied magnetic field as reported by McGuire *et al*[39]), and a Curie temperature of 54 K. The observed reduction in the magnetic moment in the powder samples can be attributed to several factors such as the degradation of the $CrI_3$ powder which is extremely sensitive to moisture and light[40] (despite our efforts to handle the sample in an inert atmosphere), the random orientation of microcrystals to the applied magnetic field, and the presence of both antiferromagnetic (AFM) and ferromagnetic (FM) interactions between different layers. In contrast, the M(T) curve for the 1D heterostructures has no obvious trace of a phase transition, together with a reduced magnetic moment per Cr atom for the 1D heterostructures from the M (H) curves compared to the powder. In addition, we conducted mean-field calculations of the magnetization for the tubes at various temperatures between 0 and 100 K, observing a distinct indication of an ordering transition around T ~ 60 K for all tube diameters examined (see more details in SI, **Figure S7**). Although the reduced moment is compatible with the non-collinear radial magnetic state predicted by theory, our sample comprises an ensemble of tubes with different lengths and orientations. However, this behavior can be described by the magnetic behavior of inhomogeneous systems as reported by a recent work[41] where magnetic nanoparticles can be strongly affected by local structures, highlighting the need for local probes to fully characterize the magnetic state of the individual nanostructures.

To directly probe the electronic and magnetic behavior of these samples, we have employed element-specific XAS and XMCD spectroscopies at the BOREAS beamline of ALBA synchrotron.[34,35,42] Field-dependent as well as temperature-dependent XMCD measurements across the Cr $L_{3,2}$ and I $M_{2,3}$ and $M_{4,5}$ edges were performed on both reference the $CrI_3$ powder and the $CrI_3$ 1D vdW heterostructures, as depicted in **Figures 4**, **S8** and **S9**.

The XAS spectra shown in **Figures 4**, **S8** and **S9**, displays Cr *L* edges and Iodine *M* edges both for the powder and the 1D nanotube. The very large dichroism in the XMCD Cr $L_{3,2}$ spectra found in the powder samples (**Figures 4a** and **S9a**) clearly evidences for a Cr-localized magnetic moment, both at a 6T magnetic field and in the remanent state (nominal $\mu_0H=0$) field. A still large XMCD signal for the $CrI_3$ 1D vdW heterostructures sample (**Figure 4b**) is also observed under a 6T applied field, which unequivocally corroborates the existence of sizable magnetic moments in the nanotubes associated with the Cr atoms. The corresponding remanent state (**Figure 9b**) shows very tiny and noisy Cr XMCD, indicating a very small magnetic remanence of the $CrI_3$ 1D vdW heterostructures sample. The temperature dependence of the XMCD, (both at zero field) notably, the XAS line shape of the $CrI_3$ 1D vdW heterostructure presents a marked difference with respect to $CrI_3$ reference powder both in the multiplet structure as well as in the energy position of the most intense features, possibly reflecting the different crystal field on account of the curvature and the proximity to the carbon layer.

Employing the sum rules analysis to the Cr XMCD spectra, measured in the highest available magnetic field of 6 T and at T=4 K, we calculated the local effective spin $M_S$ and orbital $M_L$ magnetic moments for the $CrI_3$ reference powder and $CrI_3$ SWNT@MWCNT, which yields $M_S=1.52\pm0.06$ $\mu_B$, $M_L=-0.01\pm0.02$ $\mu_B$ and $M_S=0.36\pm0.05$ $\mu_B$, $M_L=0.01\pm0.02$ $\mu_B$ respectively (for details on the calculations see **Figure S10**). The expected spin moment for a $Cr^{3+}$ atom is close to $3\mu_B$[30], thus, the magnetic moments of both the $CrI_3$ powder and the $CrI_3$ nanotubes are smaller than those reported for $CrI_3$ crystals. The discrepancy observed on the $CrI_3$ powder, seen both with SQUID and XMCD, can be attributed to a lack of magnetic saturation at 6 T as a result of the random orientation of the $CrI_3$ microcrystals easy axis with respect to the external magnetic field, and to a small spurious Cr XAS component arising from a fraction of non-magnetic oxidized powder. The lack of magnetic saturation is even more striking in the $CrI_3$ 1D vdW heterostructures sample, as the spin effective moment is about 8 times lower than the bulk value, indicating a different magnetic behavior of our 1D heterostructures. In order to understand the magnetic behavior of $CrI_3$ nanotubes, particularly the test of the prediction of radial magnetization, it is crucial to conduct single-tube experiments. Such experiments would provide direct evidence of the magnetic ordering and help confirm whether the observed reduction in magnetic moment is due to radial magnetization or another effect, such as tube orientation disorder. Our initial results are consistent with a radial magnetic state which is discussed in the following section, but other scenarios, such as tubular geometry inducing non-collinear magnetization, increased anisotropy, or multidomain states, cannot be ruled out.

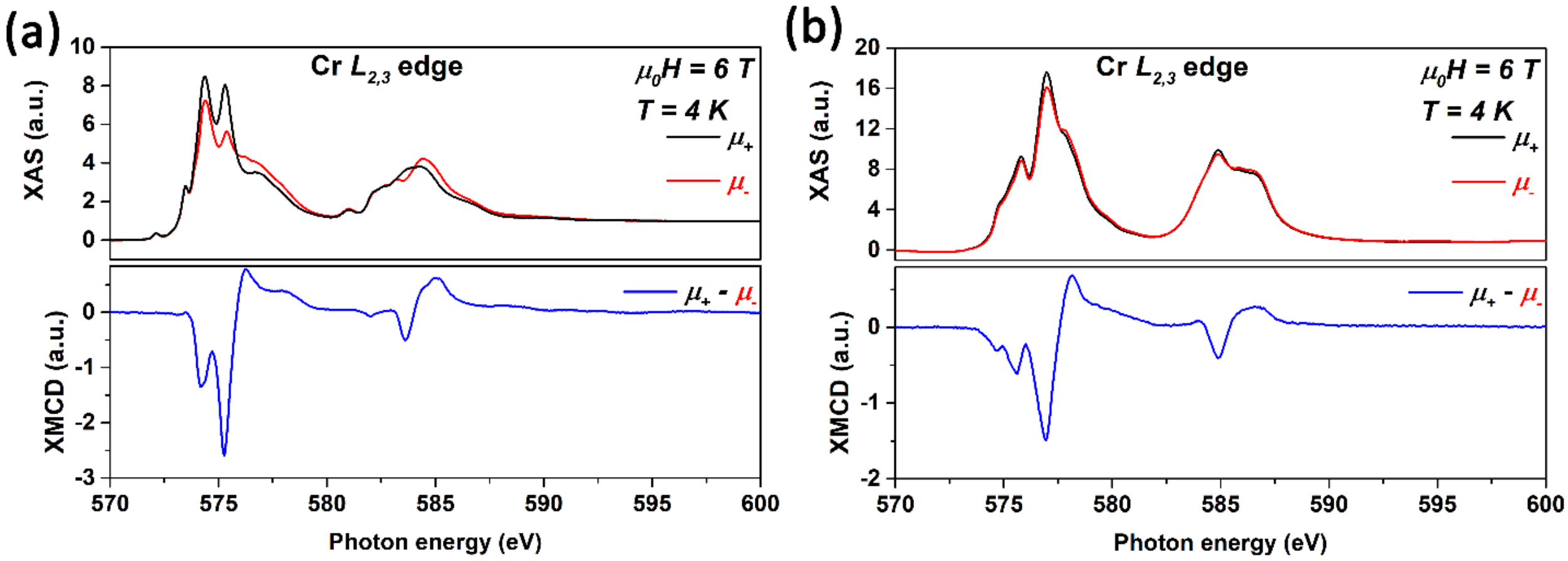

**Figure 4**. Magnetic properties of the samples: XAS with positive and negative circularly polarized light and XMCD taken at the Cr $L_{2,3}$ edge in a field of 6 T at 4 K temperature for (a) $CrI_3$ powder and (b) $CrI_3$ NT@MWCNT.

## 2.1. Theory

Compared to monolayer $CrI_3$, our structures have three main differences: curvature, dimensionality, and proximity to a carbon layer. The electronic and magnetic properties of the synthesized $CrI_3$ nanotubes were analyzed using two different levels of theory. First, density functional theory (DFT) calculations were used to assess the basic electronic and magnetic properties of one- and two-dimensional heterostructures of $CrI_3$ with carbon nanotubes and graphene, respectively (**Figures S12** and **S13**). The calculated one-dimensional structures include $CrI_3$ ribbons and tubes, both free-standing (**Figure S14** and **S16**) and encapsulated in carbon nanotubes (CNT) (**Figure S15**). Second, a model Hamiltonian describes how the competition between radial anisotropy and exchange determines the emergence of radial magnetization in tubes.

Our preliminary study of free-standing nanotubes and nanoribbons of $CrI_3$ with different chiralities and edge constructions (**Figure S15**) analyzed their relative thermodynamic stability as a function of size. Both our calculations and previous work[43] demonstrate that ferromagnetic ordering remains more stable than antiferromagnetic one, in all tolerably stressed $CrI_3$ nanostructures. The stabilities of $CrI_3$ nanotubes and nanoribbons are governed by radius R as ~$1/R^2$ and ~$1/R$ functions[43,44] crossing at R = 14 Å as the critical point where single-walled $CrI_3$ nanotubes become thermodynamically more stable than nanoribbons, and their mass production is not hindered by the elastic strain of $CrI_3$ layers (**Figure S15**). All studied $CrI_3$ nanostructures are semiconductors, where the top valence band consists of a mixture of I5*p*- and Cr3*d*-states and the bottom conduction band consists of Cr3*d*-states (**Figures S18-S20**).

We then carried out DFT calculations for $CrI_3$ tubes and ribbons encapsulated in CNT (see **Figure 5** as well as **Figure S17**). Our model set was limited to "extreme" variants of layers' curvature existing in either a narrow coaxial $CrI_3$ nanotube/nanoribbon and carbon nanotube ($CrI_3$NT@CNT or $CrI_3$NR@CNT) or the planar heterostructures of a $CrI_3$ monolayer and graphene ($CrI_3$||C) as a prototype of large $CrI_3$@CNTs (**Figure S18**). The results show the preservation of ferromagnetic state for $CrI_3$ within all heterostructures irrespective of both the morphology of $CrI_3$ and the conductivity of carbon, yet often with a weakening of spin coupling on 2-25% (**Table S2**). The main purpose of our computational study is to determine the structural stability of the tubes, the survival of the magnetic moments, and the charge transfer between MWCNT and $CrI_3$. As a byproduct, our calculations also provide an estimate of J, but we have not made a systematic study of how such a prediction would depend on the values of Hubbard U in the calculations.[45]

Our band structures of both $CrI_3$@CNT and $CrI_3$/graphene heterostructures reveal the alignment of the conduction band of $CrI_3$ with the Dirac point of carbon. In line with previous work[46], we find a spin-dependent hybridization of graphene Dirac cones with the conduction band of $CrI_3$. We also find the same phenomenon in the tubes (see, for instance, (3,3) $CrI_3$ NT, **Figure S18**). In both cases, our calculations find charge transfer is in the range of 5 $10^{12}$ $cm^{-2}$, in line with the charge transfer obtained for 2D $CrI_3$/graphene interfaces (**Figure S18**, **Table S3**). Both our DFT calculations and previous work[24,44,47] show that isolated $CrI_3$ tubes remain insulating down to very small radii. Long-range magnetic order is normally suppressed in most one-dimensional models. However, the spin-correlation lengths could easily be longer than the length of the tubes, gapping out long-wavelength fluctuations so that the tubes behave like nanomagnets. Our results on the electronic structure of $CrI_3$||C heterostructures obtained using DFT calculations with a strictly localized basis set are in line with the results from numerous DFT studies using only a plane-wave basis set.[14,16,48,49] The charge transfer estimated by using a parallel plate capacitor model is considered as significant enough to affect the work function of $CrI_3$||C[46] is in the same range as the results obtained here. Based on the similar magnetic properties of $CrI_3$/graphene two-dimensional structures[46], where the charge transfer does not affect the magnetic properties of $CrI_3$, we expect that the properties of the $CrI_3$ tubes inside the MWCNT will be very similar to those of isolated $CrI_3$, that have been modeled recently.[24] These DFT calculations[24] predict a radial magnetic state in *armchair* nanotubes when the perimeter is above a threshold of 9 nm that our *zigzag* tubes surpass.

The radial state is the natural consequence of the off-plane magnetic anisotropy of $CrI_3$ monolayers. The radial state will minimize the magnetic anisotropy energy but it implies an overhead in the exchange energy, as the spins are no longer collinear. This can be captured by the following micromagnetic model for a honeycomb lattice on a tube, with the zigzag rows along the azimuthal direction (see **Figure 5**) so that every atom in the tube has two first neighbors with different azimuthal angle and one more atom with the same azimuthal angle. We assume first-neighbor ferromagnetic isotropic exchange and single-ion anisotropy along the off-plane (radial) direction. The intracell free energy per unit cell is

$$U \cong -J \sum_{i=1,N} \vec{M}_i \cdot \vec{M}_{i+1} - D \sum_{i=1,N} \left(\vec{M}_i \cdot n_i\right)^2$$

Where J and D are positive constants, $\vec{n}_i = (cos\theta_i, sin\theta_i, 0)$ are unit vectors pointing along the radial direction, and $\vec{M}_i$ stand for the magnetization direction vector, with norm $S = 3/2$. We consider two magnetic states, collinear and radial. They can be expressed as $\vec{M}_i = S(1{,}0{,}0)$, for the collinear case, and $\vec{M}_i = S\vec{n}_i = S(cos\theta_i, sin\theta_i, 0)$ for the radial case.

The collinear state minimizes the exchange energy, whereas the radial state minimizes the anisotropy energy. We note that the inter-cell exchange interaction will give the same contribution for both states. As a result, the total energy per unit cell reads, for the collinear and radial configurations:

$U \cong -NS^2 \left(J + \frac{D}{2}\right)$ (collinear)

$U \cong -NS^2 \left(J(1 - \frac{(\Delta\theta)^2}{2} + D\right)$ (radial)

Where $\Delta\theta = \frac{2\pi}{N}$ and we have approximated $\sum_{i=1}^{N} cos^2\theta_i \simeq \frac{N}{2}$, and also use the Taylor expansion for $\cos\Delta\theta$ in the second line. From these equations, we see that, in the limit of a large perimeter, the radial state is more stable than the collinear state. We can estimate the critical perimeter, in terms of N, above which the radial state is stable, using $(\Delta\theta)^2 = \left(\frac{2\pi}{N}\right)^2 = \frac{D}{J}$.

We now take $D = 0.13$ meV, which gives a magnetic anisotropy energy of 0.4 meV, and $J = 2.2$ meV[15], and we find $\Delta\theta = \sqrt{\frac{0.13}{2.2}}$ from which we obtain $N = 26$. For a zigzag nanotube with Cr-Cr distance of 3.87 Å, this corresponds to a diameter of 3.15 nm, much smaller than most of our tubes. Therefore, based both on this analysis and on DFT calculations[24] we conclude that the minimal energy magnetic configuration is a radial magnetization state, which is non-collinear and presents a zero-net macroscopic moment. The stability of this state with respect to quantum fluctuations, as well as the impact of long-range dipolar interactions, remains to be addressed. Such a radial magnetization state could have a non-trivial impact on the electronic transport properties of the graphene nanotube. The non-collinear spin texture will be felt by the electrons in graphene as an effective spin-dependent potential that would resemble a spin-orbit coupling[50], which can substantially alter the electronic structure of the nanotube and thus affect electronic transport.

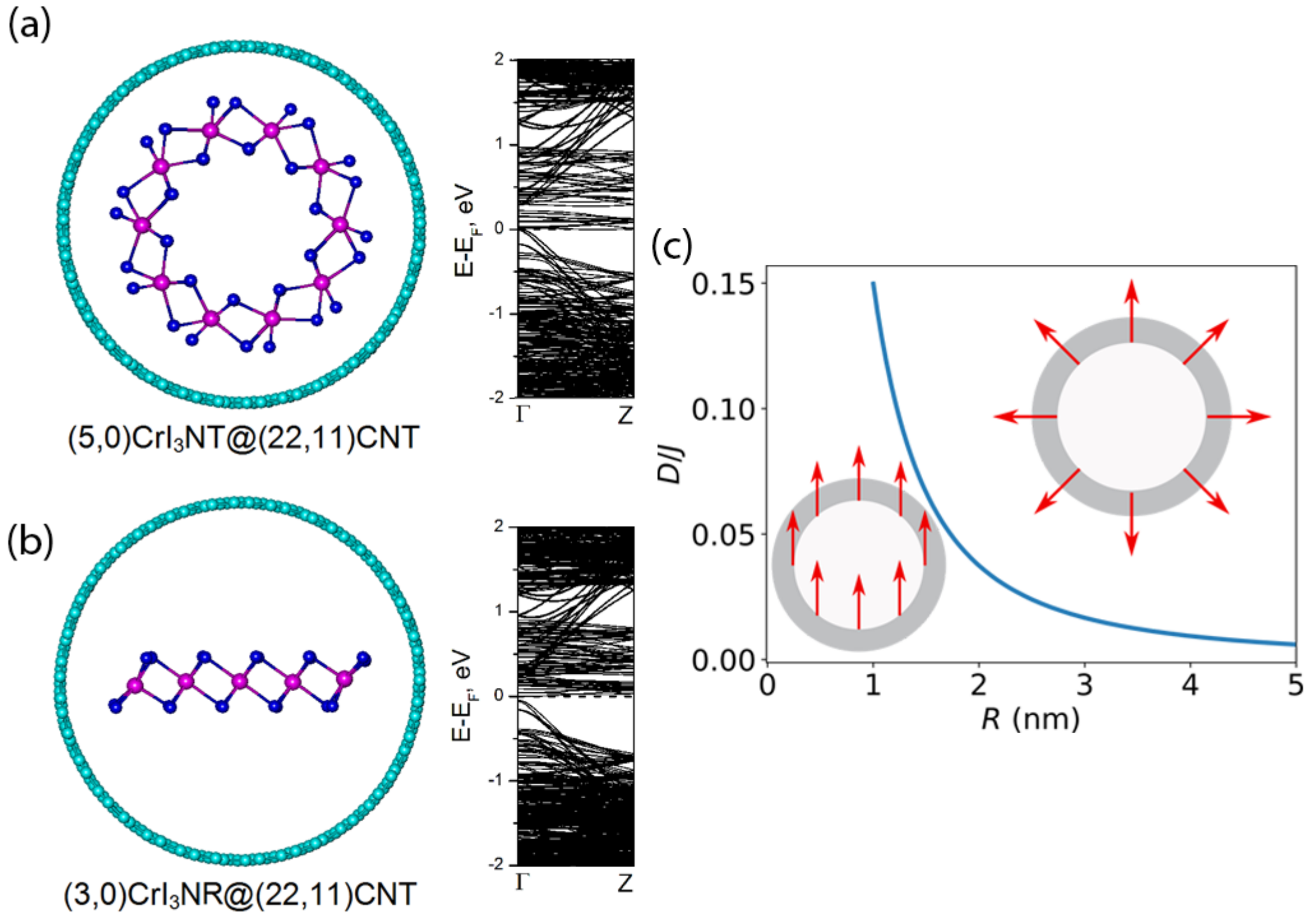

**Figure 5**. (a) Axial projection of the model 1D heterostructures assembled of semiconducting carbon nanotube (CNT) and $CrI_3$ nanotube (NT) and (b) nanoribbon (NR) together with their band structures calculated using DFT method including spin-orbit interactions: All heterostructures behave as semiconducting heterojunctions of type II, where the bottom and top edges of the bandgap arise from the valence band of carbon and the conduction band of $CrI_3$, respectively. (c) Ground state magnetization: Schematic representation of the phase diagram for the lowest energy magnetization configuration of $CrI_3$ nanotubes, showing the regions in the (radius, anisotropy: exchange) plane where collinear and radial configurations are energetically more stable. The blue line represents the boundary between the two regions which, according to our model calculations, has the form $(D/J)\sim 1/R^2$ (see discussion in the main text).

## 3. Conclusions

In conclusion, we successfully obtained 1D $CrI_3$ magnetic single-walled nanotubes (encapsulated within MWCNT nanotubes) by using the chemical vapor transport method. High-resolution electron microscopy results revealed that the 1D $CrI_3$ structures crystallized with zigzag chirality within the host nanotube system. The host MWCNT nanotube vessel stabilizes and protects the $CrI_3$ materials, allowing the formation of both 1D single-walled nanotubes. Element-specific XMCD established a significant magnetic moment for chromium atoms in $CrI_3$ nanotubes under moderate to large applied magnetic fields. These results represent a significant step forward in experimentally establishing the non-trivial geometry of 1D van der Waals heterostructures as a playground for the exploration of unprecedented non-collinear magnetism and emerging properties arising from the interplay between magnetic anisotropy and curvature in tubular geometries at the nanoscale. We believe that these new 1D van der Waals heterostructures, combining 1D structures made of an insulating magnetic compound, wrapped with a conducting nanotube, can provide the realization of a nanoscale electromagnet and will spark a new field of study, where new physical phenomena can be explored. An obvious next step in this direction includes magneto-transport experiments in single $CrI_3$/carbon nanotube devices.

## 4. Experimental Section/Methods

*Synthesis of the 1D vdWs Heterostructures*: MWCNTs (without the presence of catalysts) were purchased from Nanocyl SA, Belgium, and US Research Nanomaterial, Inc., USA. The filling material was Chromium (III) iodide, with a 99.99% purity procured from American Elements (CAS # 13569-75-0). Growth of the 1D $CrI_3$ vdW heterostructures using Carbon Nanotube Templates: The synthesis of single walled $CrI_3$ magnetic nanotubes was carried out by the capillary filling method using MWCNTs as host templates. This procedure has been previously successfully employed to fabricate $BiI_3$ single walled nanotubes encapsulated within CNTs[29,30]. The CNT sample was purchased from commercial suppliers and used as-received (open-ended nanotubes devoid of any metal impurities or contaminants). The metal halide ($CrI_3$) used in this report has a low melting point (> 600 $^{o}$C) and is hygroscopic. In addition, it is extremely sensitive to light, moisture, and ambient conditions. Hence special care was taken to carry out the reactions. The nanotubes are filled using the following procedure: first, the CNT and $CrI_3$ powders are mixed, in different weight ratios (1:30, 1:20, 1:10, and 1:5, respectively), inside an Ar-filled glovebox. Among the studied weight ratios, 1:10 was the best according to their filling yield ratio as confirmed by HAADF-STEM, and this sample was used for the rest of the experiments. This mixture is packed into a quartz ampoule of length 200 mm sealed under a vacuum (below 1 Pa) and placed inside a two-zone horizontal tube furnace. Depending on the metal halides' melting point (mpt), the procedure involved heating and cooling above and below the halide's mpt for several hours to ensure the formation of the filled phases. vdWHs of single-walled $CrI_3$ nanotubes encapsulated with CNTs were successfully achieved by this method. Since capillary forces drive the metal halide melt to the tube's interior, where it solidifies upon cooling, the cycling was carried out with the dwell temperature some tens of degrees (here, ±50 °C) above and below the melting point to optimize the filling yield. The heating process followed the profile shown in **Figure S1**. On completion, the furnace is cooled naturally down to room temperature; the sealed ampoule is opened inside an Ar-filled glovebox, and the products are collected for further analysis.

*Characterization*: For TEM/STEM/4D STEM/EDS studies, fresh samples were prepared by first dispersing the powder in ethanol and then drop-casting this onto a holey carbon film 300 mesh Cu grids under inert processing conditions. A FEI Titan Themis 60-300 kV electron microscope, operated at 200 kV, equipped with probe and image correctors and a monochromator, was used for the TEM/4D STEM/STEM imaging. The EDS spectra were obtained using a Super-X detector and crystal models were built using Crystal Maker software. The 4D-STEM was employed at 200 kV with a convergence semi-angle of 21 mrad and an approximate beam current of 30 pA. A pixelated detector, the ARINA from DECTRIS, was used to capture a diffraction pattern at each probe position. This camera has a resolution of 192 × 192 pixels and operates at a frame rate of 120,000 Hz. The full dataset consists of 512 × 512 probe locations and is processed using the Python-based py4DSTEM[51] module. Raman scattering measurements were carried out with an alpha300 R confocal Raman microscope (WITec) using a 532 nm and 633 nm Nd:YAG laser for excitation. The system was operated with an output laser power of 1.3 mW. The laser beam was focused onto the sample by a 50x objective (Zeiss) and the spectra were collected with 1800 groove/mm grating using 5 acquisitions with a 2 s acquisition time. Before employing magnetization measurements, the mixture of the powder was washed to eliminate the magnetic responses from the $CrI_3$ powder outside the CNTs. The mixture of the powder was constantly stirred in distilled water for 24 h and then filtered with a filter paper. A Superconducting Quantum Interference Device (SQUID) magnetometer (Quantum Design, SQUID-VSM) was used to analyze the magnetic properties of the filled $CrI_3$ nanotubes by weighing and fixing them with cotton into a gelatin capsule and then placed in a straw sample holder. The magnetization behavior of the CNT-filled $CrI_3$ nanotubes was probed by soft X-ray absorption (XAS) and X-ray magnetic circular dichroism (XMCD) at the BOREAS beamline of the ALBA synchrotron in Spain.

*DFT Calculations*: The spin-polarized density-functional theory (DFT) calculations of all $CrI_3$ and $CrI_3$-C nanostructures were performed under periodic boundary conditions as implemented in the SIESTA 4.1.5 code[52,53]. The Perdew-Burke-Ernzerhof (GGA PBE) parametrization of the exchange-correlation potential was used. The core electrons were frozen, applying norm-conserving Troullier–Martins pseudopotentials. The wave functions of all valence states were described using the double-ζ polarized basis set. The model single-walled $CrI_3$ nanotubes can be folded using the same construction principles as for nanotubes of carbon[54] or other hexagonal compounds[55]. The calculations of the unit cells were performed with the global geometry optimization. The use of the simplified rotationally invariant DFT+U formalism[56] demonstrated only a marginal influence of the Coulomb repulsion on both the value and the edge composition of the fundamental bandgap, thereafter, $U_{eff}$ was not used. The spin-orbit coupling within all atoms, when required, was treated by means of an on-site approximation for the spin-orbit matrix elements[57]. More technical details are provided in the Supporting Information.

**Supporting Information**

Supporting Information is available from the Wiley Online Library or from the author.

**Acknowledgements**

We acknowledge financial support from the European Union (Grant FUNLAYERS - 101079184), Portugal FCT (Grant No. PTDC/FIS-MAC/2045/2021), the Swiss Science National Foundation Sinergia (Grant PIMAG, CRSII5_205987), Advanced Materials program by MCIN with funding from European Union NextGenerationEU (PRTR-C17.I1) and by Generalitat Valenciana (MFA/2022/045). Facilities for DFT computations were granted by the Institute of Solid State Chemistry UB RAS (grant No. 124020600024-5). M.V. Acknowledges funding by grant PID2020-116181RB-C32 (AEI/FEDER). The authors would like to thank Mr. James Peters and Dr. Fatima Cerqueira for their help with the Raman measurements and data analysis.

Received: ((will be filled in by the editorial staff))

Revised: ((will be filled in by the editorial staff))

Published online: ((will be filled in by the editorial staff))

# Supporting Information

## Single-walled $CrI_3$ nanotubes encapsulated within multiwall Carbon Nanotubes

[1]*Ihsan Çaha*, [1]*Aqrab ul Ahmad*, [1]*Loukya Boddapatti*, [1]*Manuel Banõbre*, [1]*António T. Costa*, [2]*Andrey N. Enyashin*, [3]*Weibin Li*, [3]*Pierluigi Gargiani*, [3]*Manuel Valvidares*, [1]*Joaquín Fernández-Rossier**, [1]*Francis Leonard Deepak**

**Growth and characterization**

The heating process followed the profile shown in **Figure S1**. On completion, the furnace was cooled down naturally to room temperature; the sealed ampule was opened; and the products were collected.

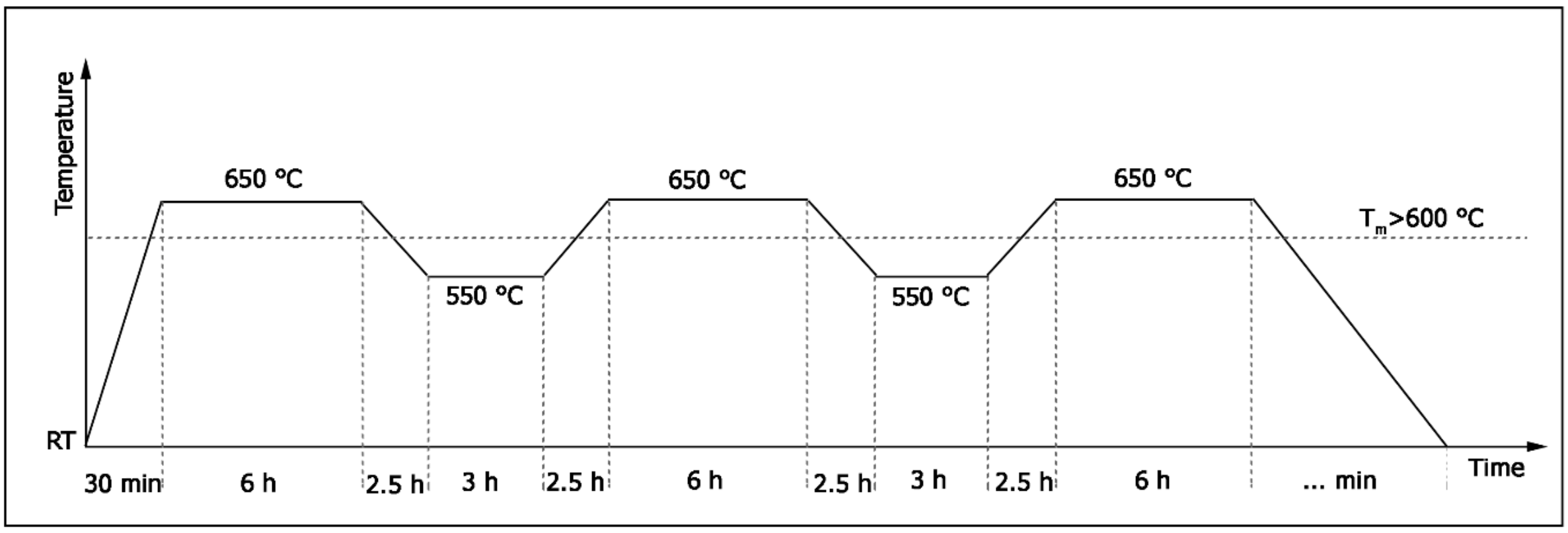

**Figure S1**. Heat treatment procedure used in the synthesis of filled CNT, where RT stands for room temperature and $T_m$ for melting temperature of the halide ($CrI_3$ as an example, mpt > 600 °C). The entire procedure typically takes around 36 hours.

HAADF and ABF-STEM images of the as-produced sample illustrate no $CrI_3$ particles outside the tubes after the washing process together with the high filling yield obtained (>80%) and the presence of long, continuously filled nanotube sections.

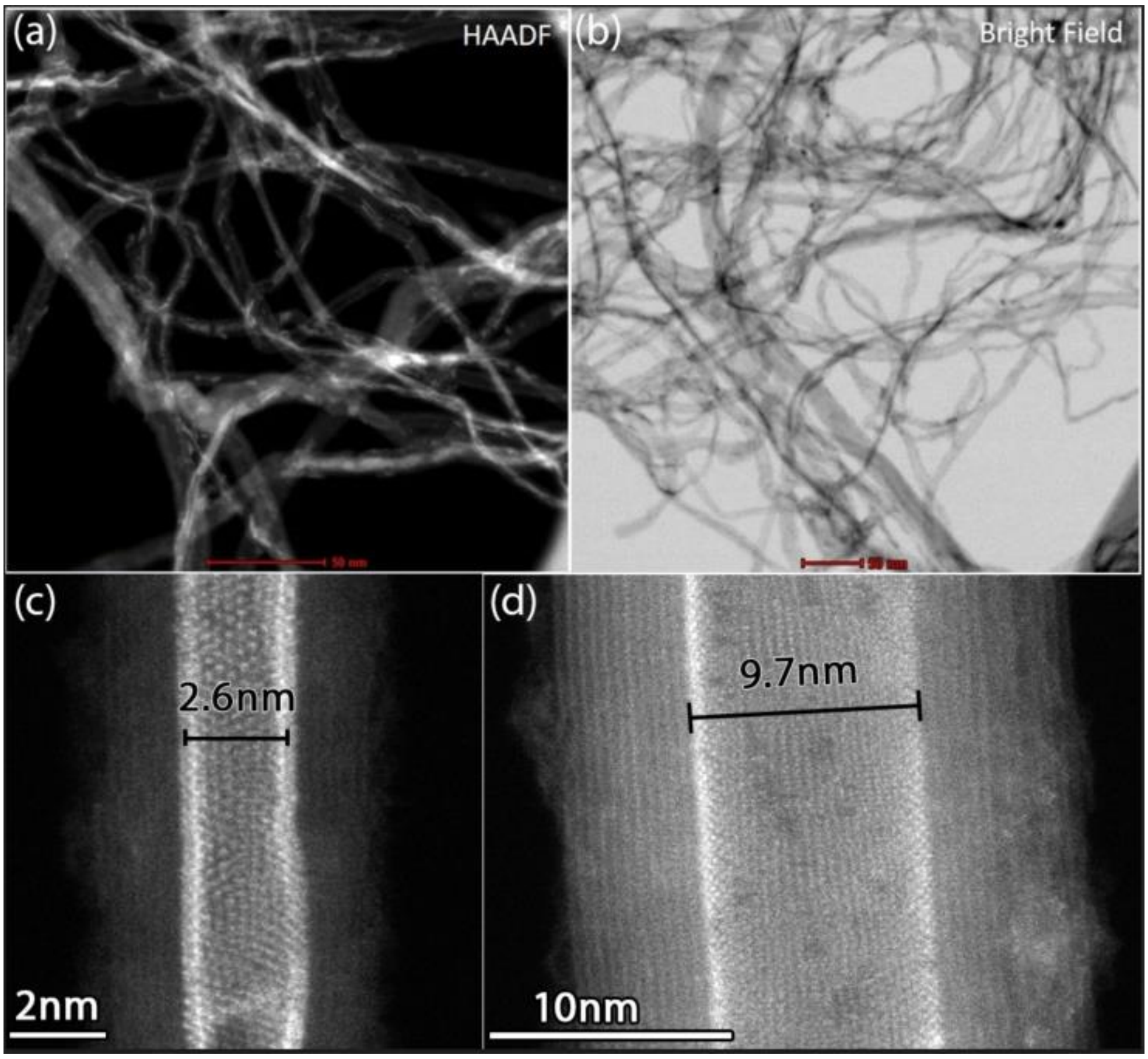

**Figure S2.** (a-b) Overview of the filled MWCNTs by HAADF and ABF-STEM images. (c-d) An example of small and large $CrI_3$ NT@MWCNTs

The strain mapping indeed shows some strain gradients, primarily due to the curvature of the nanotubes. However, these strains are smoothly distributed, as seen in both **Figure S3 (a)** and **(b)**, without the sharp discontinuities typically associated with dislocations. Despite the strain present, the high-resolution STEM images do not show clear signatures of dislocations, such as lattice disruptions or missing atomic planes. The strain appears to be elastically accommodated, indicating that the nanotube tolerates the mismatch without forming dislocations. While the strain gradient from curvature is evident, any uniaxial strain would also be smoothly distributed. The material remains in the elastic regime, where neither type of strain reaches the level necessary to induce dislocation formation. $CrI_3$, like other 2D van der Waals materials, has a high capacity for elastic deformation, which allows it to accommodate both curvature and strain without introducing dislocations.

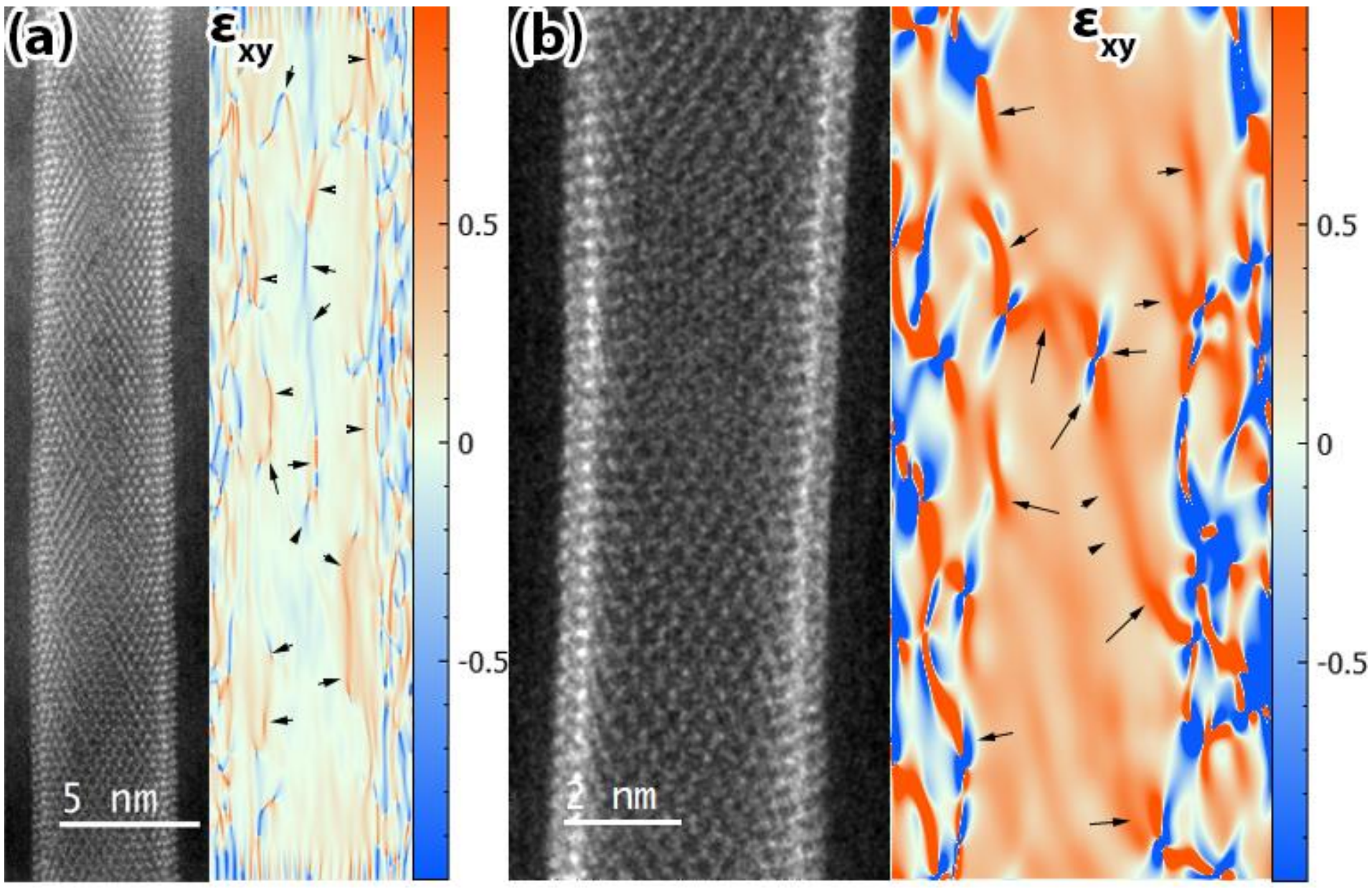

**Figure S3.** (a) and (b) HAADF-STEM image of encapsulated $CrI_3$ nanotubes and their strain tensor in xy direction obtained by GPA (Geometric phase analysis).

**Raman spectra**

The Raman spectra of a $CrI_3$-filled MWCNT sample obtained using the 532 nm and 633 nm Nd:YAG laser are given in **Figure S4** and the corresponding Raman shift positions are given in Table S1. Regarding the G-band and 2D-band, no significant shifts or shape changes were observed. This implies an absence of charge transfer between the encapsulated material and the nanotubes. However, it is important to realize that charge transfer is local: it only affects the first carbon layer of the tube and the $CrI_3$. In contrast, the Raman technique is a global probe, therefore providing information over all the carbon layers. As a rule of thumb, the MWCNTs have between 5 to 10 layers. Therefore, only 20% or less of the signal may be affected by the charge transfer.

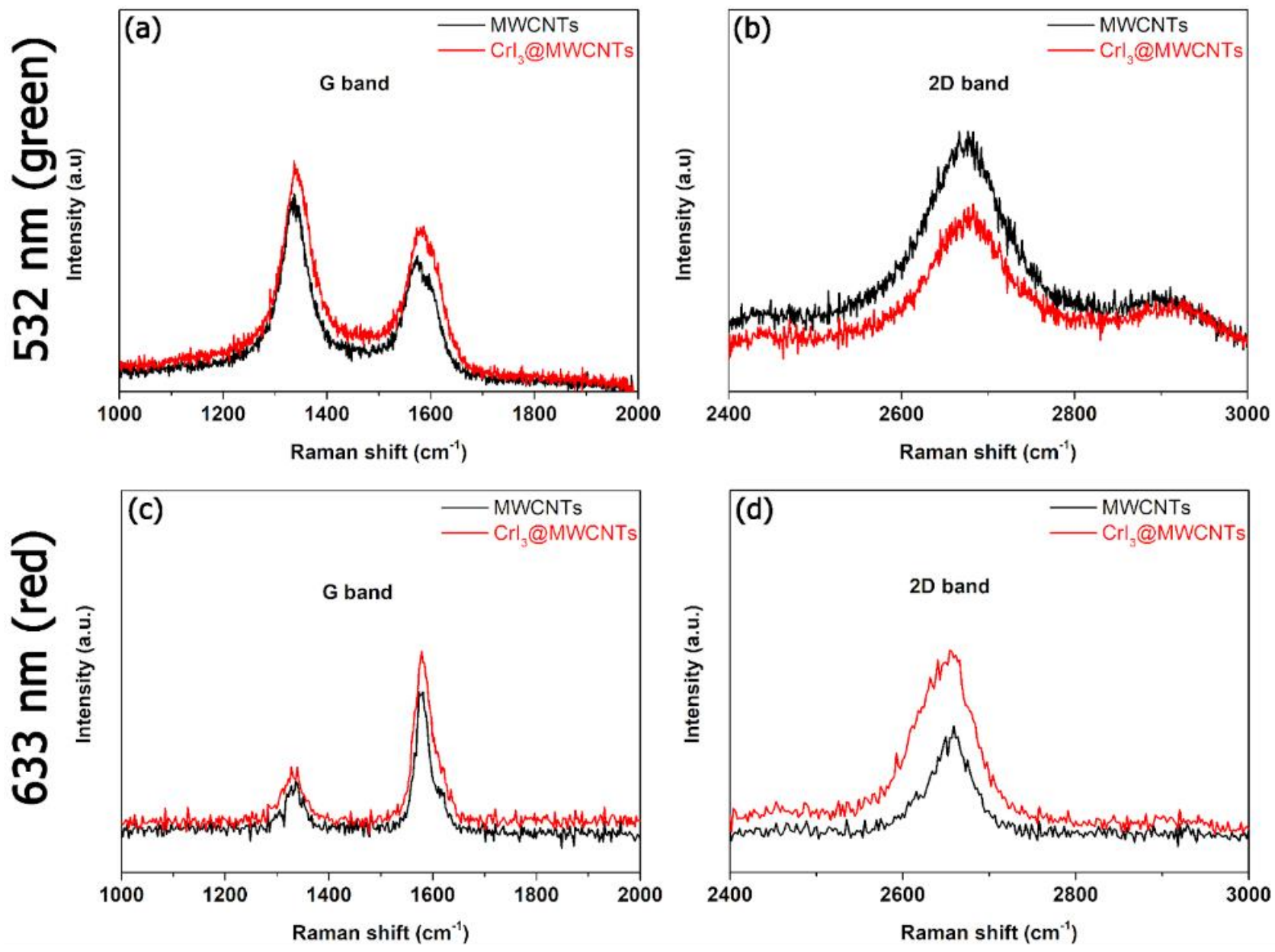

**Figure S4**. Raman spectra of the CNT and $CrI_3$@MWCNTs samples with 532 nm and 633 nm Nd:YAG laser in the two regions: (a), (c) G mode, and (b), (d) 2D mode.

**Table S1**: Raman shift positions in $cm^{-1}$ for 532 nm wavelength

| Samples | D position | G position | D* Position | 2D |
|---|---|---|---|---|
| CNTs | 1337 | 1570 | 1603 | 2675 |
| $CrI_3$@MWCNTs | 1341 | 1578 | 1608 | 2678 |
| Differences | 4 | 8 | 5 | 3 |

**4D STEM**

In recent years, sub-Ångstrom electric field and charge density mapping using 4D-STEM (Four-Dimensional Scanning Transmission Electron Microscopy) center of mass (CoM) imaging has become feasible due to advances in aberration-corrected STEM and fast pixelated detectors. To demonstrate the reduction of charge transfer through CNT walls, we employed 4D-STEM at 200 kV with a convergence semi-angle of 21 mrad and an approximate beam current of 30 pA. A pixelated detector, the ARINA from DECTRIS, was used to capture a diffraction pattern at each probe position. This camera has a resolution of 192 × 192 pixels and operates at a frame rate of 120,000 Hz. The full dataset consists of 512 × 512 probe locations and is processed using the Python-based py4DSTEM module.

To compare the experimental results, we generated a simulated HAADF-STEM image (**Figure S5c**) and simulated 4D-STEM electron charge density map of panel (c) which is a $CrI_3$ SWNT within a six-wall CNT using Dr. Probe software and custom-made Python code. **Figure S5a** presents the simulated charge density mapping, where the charge intensity decreases from the $CrI_3$ tube outward through the CNT walls. This trend is further illustrated in the line profile extracted from panel (a) (**Figure S5b**), which shows a linear decrease in charge density intensity from the CNT's inner layers to the outermost walls.

The experimental charge density mapping of a $CrI_3$ SWNT@MWCNT, corresponding to its ADF image in **Figure S5f**, is shown in **Figure S5d**, along with a line profile in **Figure S5e**. These results provide clear evidence of charge transfer from the $CrI_3$ nanotube to the first carbon layer, with a gradual decrease as it moves through the CNT walls. This finding explains why no significant charge transfer was observed in Raman analysis, despite the noticeable charge transfer predicted by DFT calculations (discussed in subsequent sections).

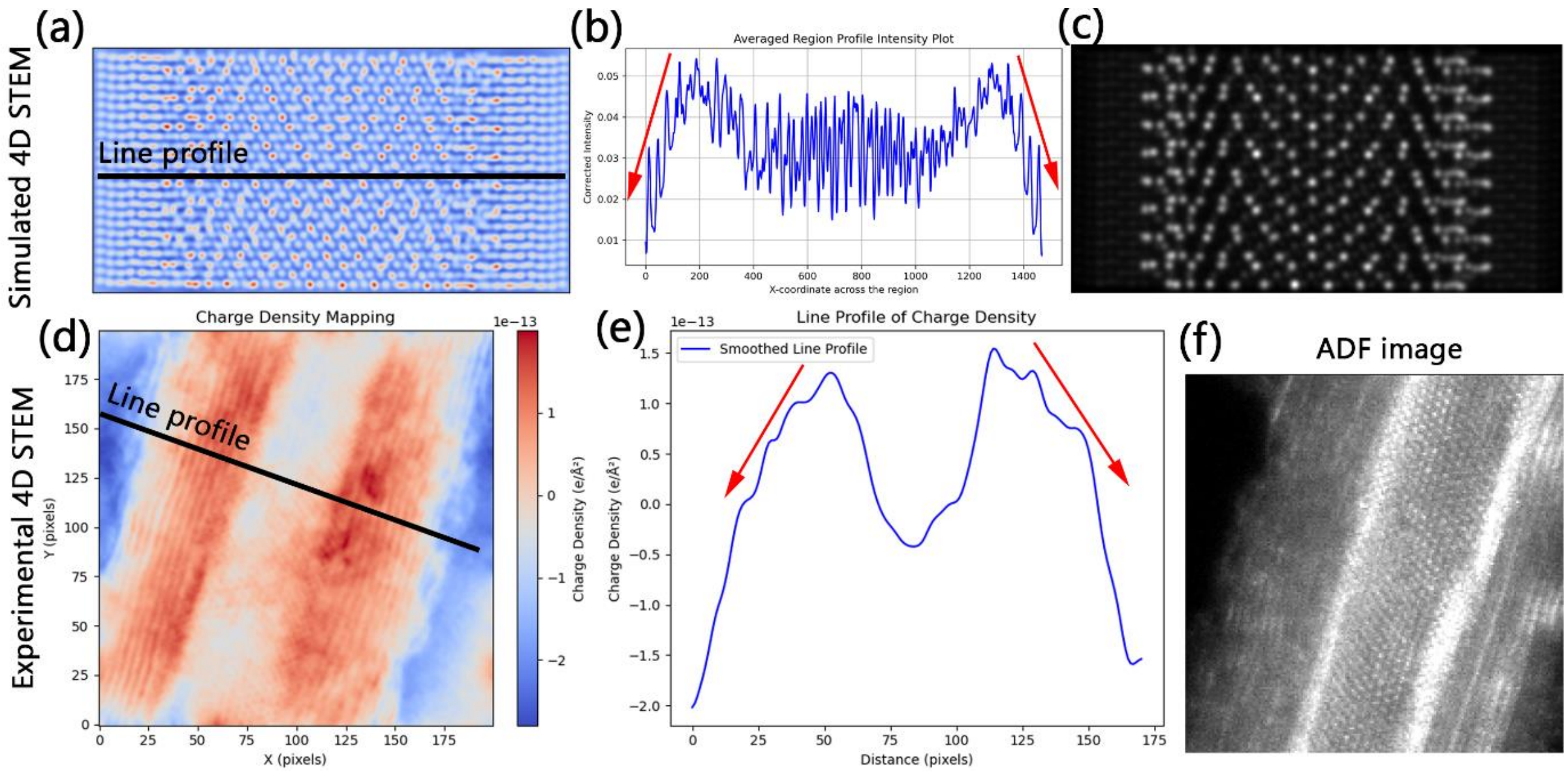

**Figure S5**. (a) Simulated electron charge density mapping. (b) Line profile intensity extracted from panel (a). (c) Simulated HAADF-STEM image. (d) Experimental electron charge density mapping. (e) Experimental line profile intensity extracted from panel (d). (f) Virtual ADF image of a $CrI_3$ SWNT@MWCNT, corresponding to the electron charge density mapping shown in panel (d).

### SQUID Measurements

**Figure S6** shows the SQUID measurements of both the $CrI_3$ powder precursor and the encapsulated $CrI_3$ tubes. **Figures S6a** and **S6b** present the temperature-dependent magnetization M(T) with zero-field cooling (ZFC) and field-cooling (FC) modes measured under 100 Oe field for the $CrI_3$ powder (S6a) and $CrI_3$ SWNT@MWCNT samples (S11b). The temperature dependence of the powder M(T) curve is consistent with 3D crystal magnetic measurements, including ferromagnetic order below a Curie temperature of 54K, slightly below the values reported for crystals, and a reduction, of a factor of roughly 2, of the magnetic moment per atom, that we attribute to the random orientation of the $CrI_3$ microcrystals easy axis with respect to the external magnetic field and to larger surface to volume ratio of powder that, together with surface degradation, may reduce the magnetization. The M(T) of the tube samples is very different, showing a monotonic decrease of M as T is increased, similar to a paramagnet. A minor feature in dM/dT at 61K is compatible with some sort of magnetic transition, affecting a fraction of the magnetic atoms. The M(H) curves for powder and nanotubes, shown in S6(c), permit us to estimate the saturation magnetic moment per Cr. For the powder, we find values around 1.8 for 2 kOe, not far from the 1.5 BM of the FC M(T) curve at a small temperature, consistent with ferromagnetic order. This value is also consistent with our sum-rule analysis from XMCD (**Figure S10**). For the nanotubes, the magnetic moment per Cr of the M(H) curve is about 0.25 BM, consistent with the value inferred from the sum rule analysis at 6 Tesla (0.36 BM). Therefore, our magnetic measurements show that the average magnetic moment of Cr is smaller in the tubes than in the $CrI_3$ powder. This is consistent with the compensation of magnetic moments averaging across different directions in an ordered radial configuration.

The isothermal magnetization versus applied field at T = 5 K for purchased CNTs, $CrI_3$ raw powder, and $CrI_3$ NT@MWCNT samples is shown in **Figure S6c**, together with the opening of the loops in **Figure S6d**. We note that the shape of the M(H) curve of the powder is different from a single crystal, with a much smaller coercive field. The purchased CNTs were diamagnetic, thus all magnetization is due to $CrI_3$ nanotubes. In contrast, the $CrI_3$ nanotubes show a larger coercive field (105 Oe), comparable to the one reported for thin-film $CrI_3$[1].

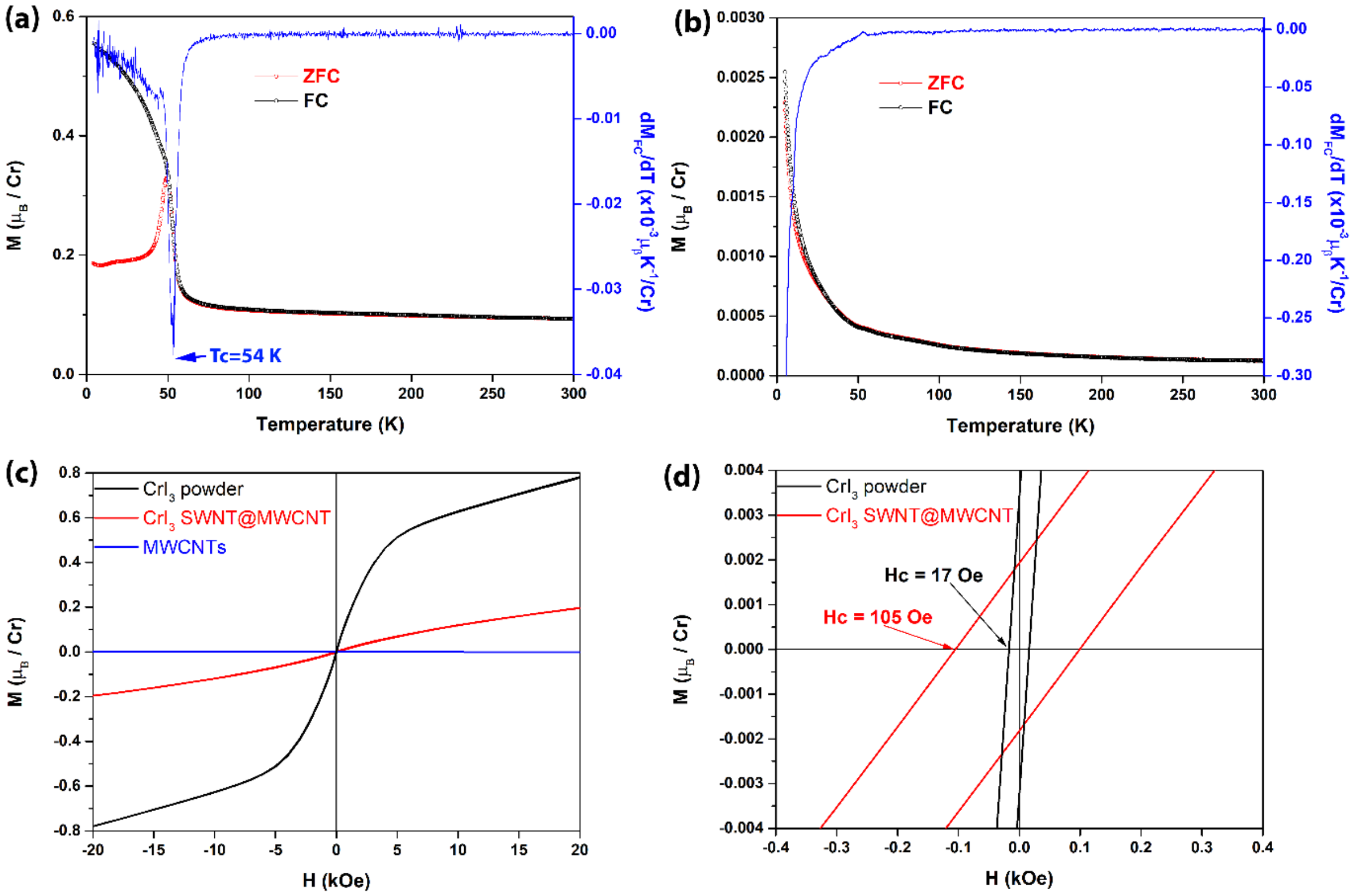

**Figure S6**. SQUID magnetometer results: M-T curve with its first derivative for $CrI_3$ powder and $CrI_3$ SW@MWCNT heterostructures in (a) and (b), respectively. FC data is taken at 100 Oe (c) M-H curve of purchased CNTs, $CrI_3$ powder and $CrI_3$ SW@MWCNT heterostructure measured at 5 K, and opening of the MH loops in (d).

**Mean field calculations for finite temperature magnetization of nanotube sample.**

We have adopted a model that contains three of the essential elements that determine the behavior of the spin degrees of freedom of the encapsulated $CrI_3$ nanotubes: Heisenberg (isotropic) exchange between neighboring spins, single-ion anisotropy perpendicular to the surface of the tube and Zeeman coupling to an external uniform magnetic field. The perpendicular anisotropy favors a radial state, competing with the Heisenberg exchange, which favors parallel alignment. For large enough tube radii, the frustration of the isotropic Heisenberg term is small enough such that the radial alignment is favored overall. The effect of the external magnetic field obviously depends on the relative orientation between the field and each tube's axis. Since the tubes have cylindrical symmetry around their respective axes, a single angle is enough to characterize this relative orientation.

We performed mean-field calculations of the magnetization of the tubes for several temperature values in the range of 0 - 100 K. To take into account the random nature of the tubes' orientations in our samples we averaged our simulation results over the direction of the external magnetic field. The values for the Heisenberg exchange (2.2 meV) and single ion perpendicular anisotropy (0.3 meV) have been taken from the literature[2]. The results are summarized in **Figure S7**. We see a clear signature of ordering transition at T ~ 60K for all tube diameters considered. By reducing the ratio between the Heisenberg exchange and perpendicular anisotropy, the transition temperature can be driven downwards, but it remains finite for every finite value of the Heisenberg exchange. Thus, we conclude that the radial order alone cannot be responsible for the negative Curie-Weiss temperature behavior observed in our samples.

We should notice that a missing ingredient in our modeling is the dipolar interaction, which may impact the results significantly. Also, according to recent work[3], the local structure of samples containing magnetic nanoparticles may drastically affect the results of experiments probing global properties, such as total magnetization vs. temperature. Our averaging procedure corresponds to assuming a uniform distribution of far-apart tubes, which completely ignores the possibility of local structures forming.

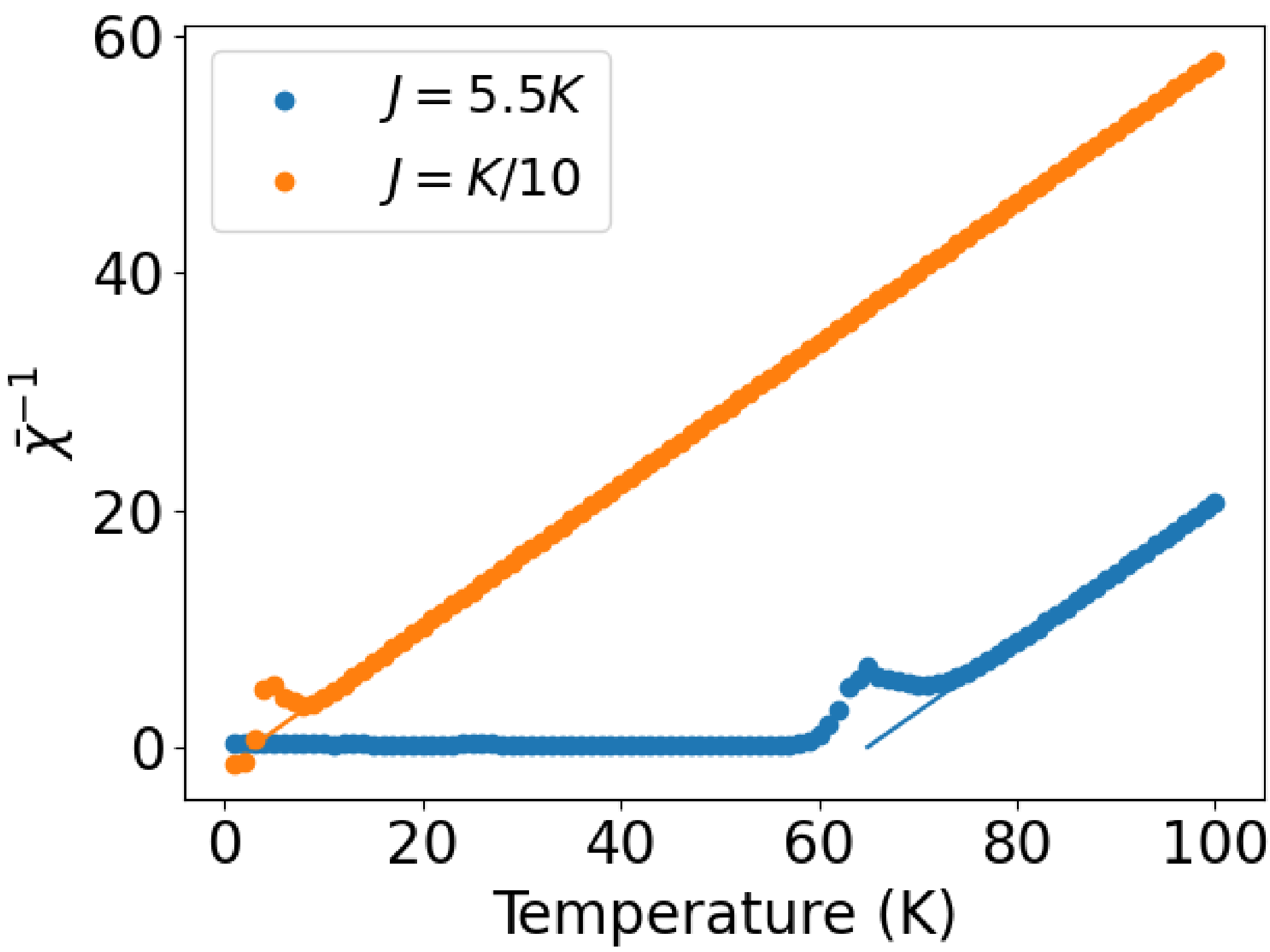

**Figure S7.** Inverse of the susceptibility associated with the component along the direction of the applied magnetic field, as predicted by a spin model containing nearest-neighbour isotropic exchange interactions and single-ion anisotropy. The blue curve corresponds to a ratio between exchange and anisotropy of 5.5, whereas for the orange curve, the exchange energy is 0.1 of the single ion anisotropy term. The perimeter of the tubes in these simulations accommodates 40 magnetic atoms. The minimal energy magnetic configuration for each individual tube at zero

external magnetic field is radial. The solid lines fit the Curie-Weiss laws, with positive temperatures of 64.9 K for *J*=5.5*K* and 2.7 K for *J*=0.1*K*.

**XAS and XMCD analysis.**

The XAS and corresponding XMCD at I $M_{4,5}$ and I $M_{3,2}$ edges for $CrI_3$ powder and $CrI_3$ NT@MWCNTs heterostructure under the 6 T field are given in **Figure S8**. Both iodine edges from XAS can clearly be seen for powder and tubes. As it is well reported in literature[4], $CrI_3$ is very sensitive and can easily degrade under ambient conditions. Therefore, we believe that this signal is coming from encapsulated $CrI_3$. Regarding magnetization at the iodine edges, there is a clear XMCD signal for powder. However, it is in the noise range or very small in the case of encapsulated tubes.

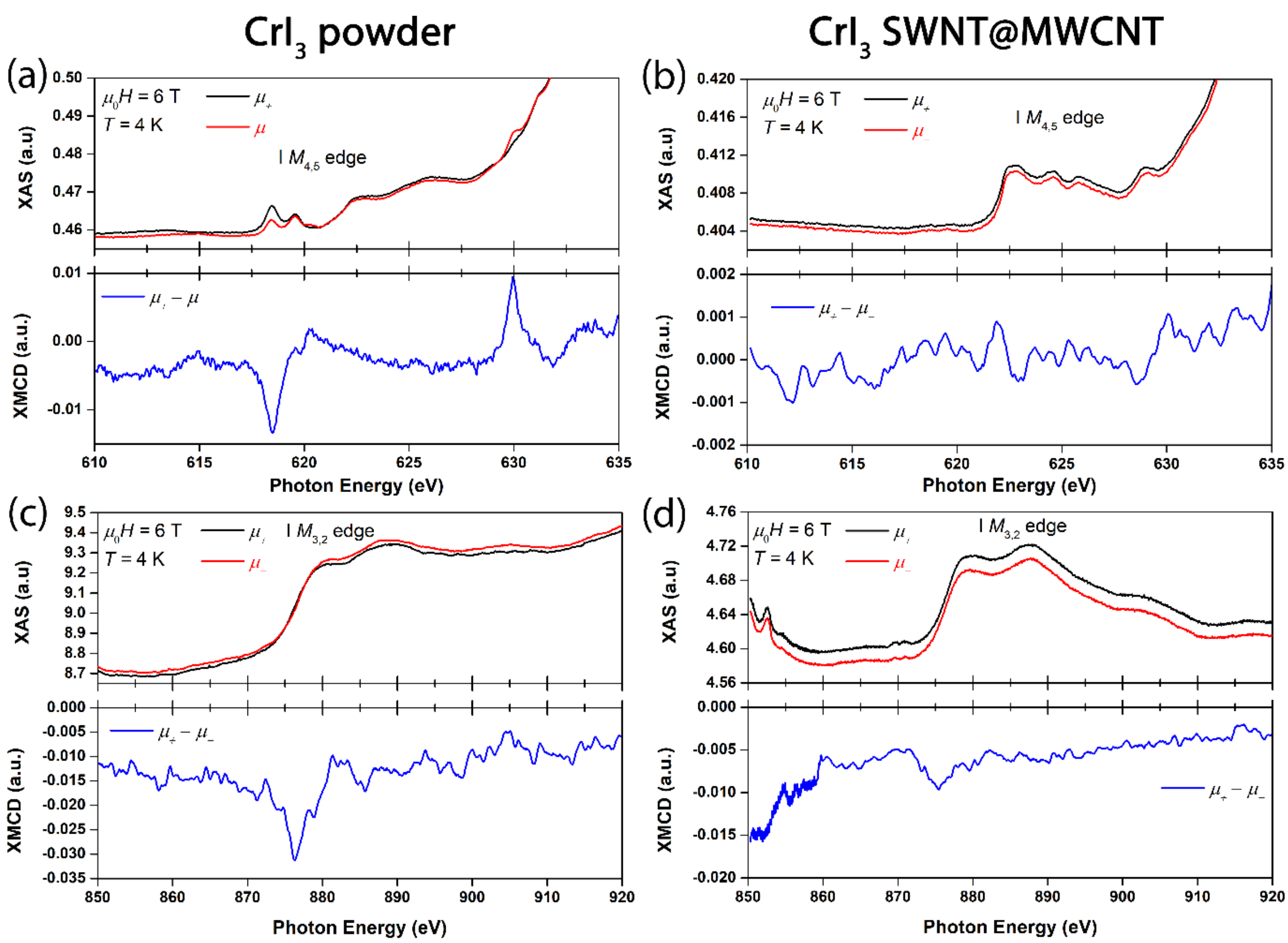

**Figure S8**. XAS and corresponding XMCD at I $M_{4,5}$ and I $M_{3,2}$ edges for $CrI_3$ powder and $CrI_3$ NT@MWCNTs heterostructure under 6 T field.

**Figure S9** presents XAS and XMCD taken at the Cr $L_{2,3}$ edges in a field of 0 T at 4 K temperature for (a) $CrI_3$ powder and (b) $CrI_3$ NT@MWCNT. Also as given in panel (c) temperature-dependent XMCD of $CrI_3$ powder and $CrI_3$ 1D vdWH taken at the Cr $L_{2,3}$ at a 6 T magnetic field showing drastically reduction of XMCD signals with increasing temperature until around 100 K.

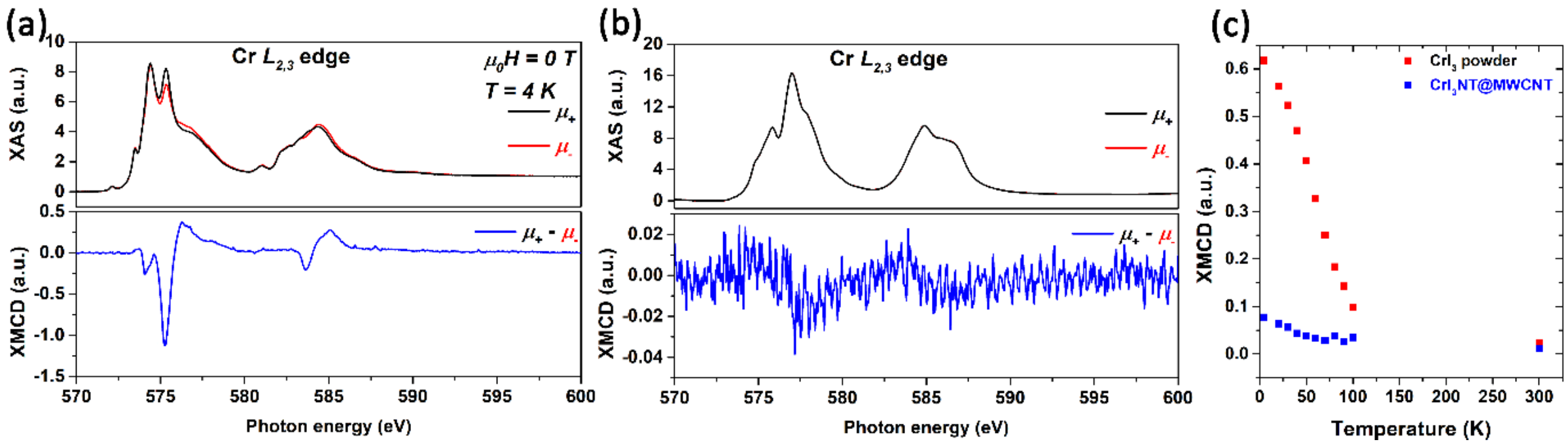

**Figure S9**. XAS and XMCD were also taken at the Cr $L_{2,3}$ edges in a field of 0 T at 4 K temperature for (a) $CrI_3$ powder and (b) $CrI_3$ NT@MWCNT. (c) Temperature-dependent XMCD of $CrI_3$ powder and $CrI_3$ 1D vdWH taken at the Cr $L_{2,3}$ edge at a 6 T magnetic field.

The XMCD sum rules plots for the $CrI_3$ powder and $CrI_3$ NT@MWCNT heterostructure at 6 T field and 4K temperature (top and bottom panels, respectively) are given in **Figure S10**. Unpolarized XAS intensity (orange), XMCD (blue), XMCD integral (green), XAS non-resonant background (black dotted line), XAS integral after background subtraction (red). For the sum rules analysis it has been assumed a number of 3d holes of 7 corresponding to a nominal $Cr^{3+}$ valence state. For the effective spin moment calculation, it has been included a 2.19 correction factor to be taken into account for the $L_3$ - $L_2$ intensity mixing[5]. Employing the sum rules analysis to the Cr XMCD spectra, measured in the highest available magnetic field of 6 T and at T=4 K, we calculated the local effective spin $M_S$ and orbital $M_L$ magnetic moments for the $CrI_3$ reference powder and $CrI_3$ NT@MWCNT, which yields $M_S$=1.52±0.06 µB, $M_L$=-0.01±0.02 µB and $M_S$=0.36±0.05 µB, $M_L$=0.01±0.02 µB respectively (for details on the calculations see **Figure S10**). The expected spin moment for a $Cr^{3+}$ atom is close to 3µB[6], thus the magnetic moments of both the $CrI_3$ powder and the $CrI_3$ nanotubes are smaller than those reported for $CrI_3$ crystals. The discrepancy observed in the $CrI_3$ powder, seen both in SQUID and XMCD measurements, may be attributed to the random orientation of the $CrI_3$ microcrystals' easy axis with respect to the external magnetic field, the presence of a significant number of defects, and a possible contribution from non-magnetic oxidized Cr species affecting the XAS signal. In the case of the $CrI_3$ 1D vdW heterostructures, the observed reduction in the spin effective moment approximately eight times lower than the bulk value suggests a distinct magnetic behavior, likely influenced by dimensionality effects, strain, or interfacial interactions unique to the 1D geometry.

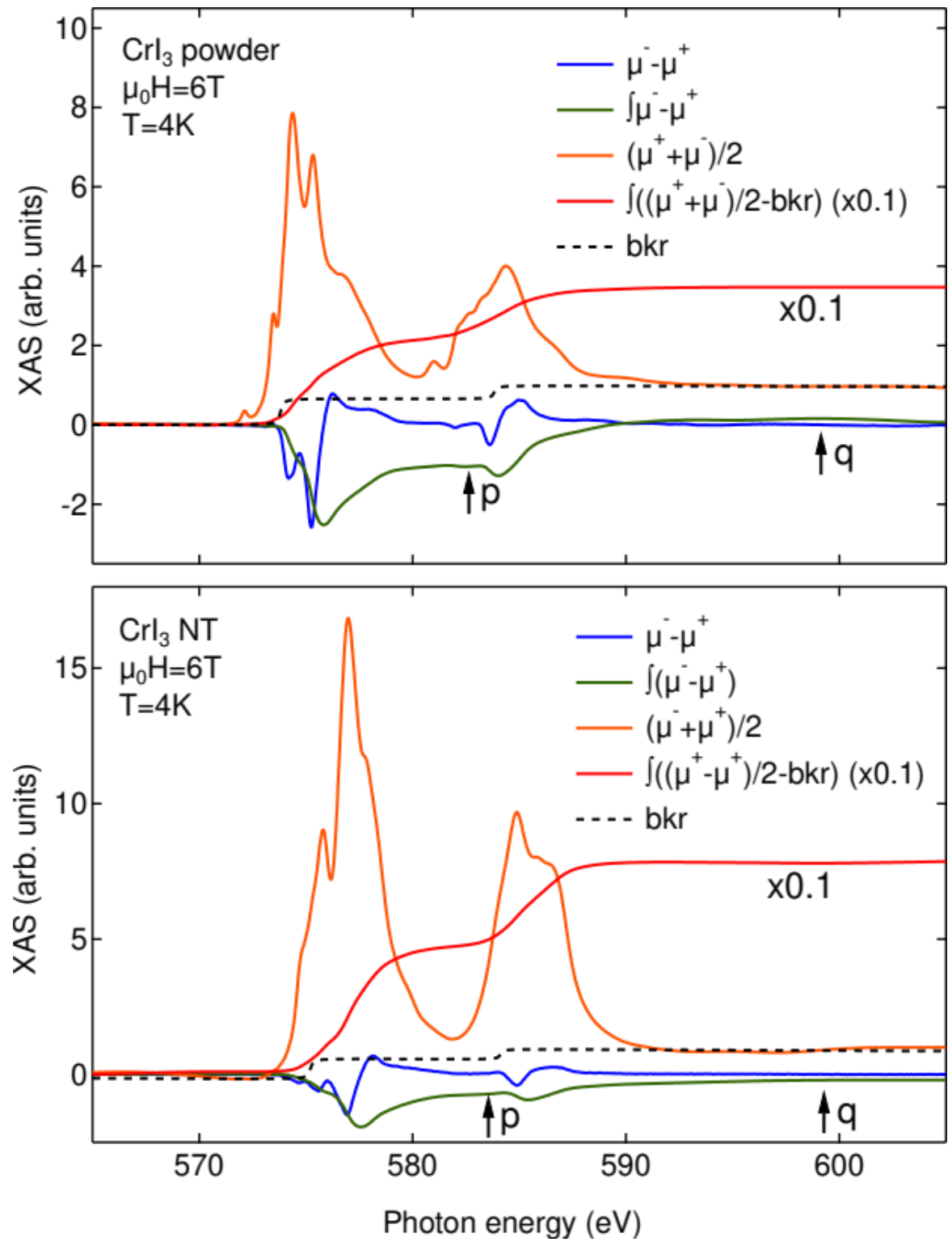

**Figure S10**. XMCD sum rules plots for the $CrI_3$ powder and $CrI_3$ SW@MWCNT heterostructure at 6 T field and 4K temperature (top and bottom panels respectively).

The XMCD hysteresis taken at the Cr $L_{2,3}$ edges at the 4 K temperature and field-dependent measurements of XMCD intensity for fields ranging from 6 T to − 6 T (T = 2 K) are given in **Figure S11**. As can be seen, the S-shape XMCD magnetic hysteresis loop in **Figure S11a** indicates the ferromagnetic phase and a vanishing signal in the paramagnetic phase. The vanishingly small hysteresis can be explained by negligible net perpendicular anisotropy resulting from the formation of different $CrI_3$ layers in CNTs[7,8]. The field-dependent XMCD result also confirmed that the paramagnetic $Cr^{3+}$ state exists and the ferromagnetic component is negligibly small because XMCD signals do not persist at B=0 T as a remanence magnetization (**Figure S9b**).

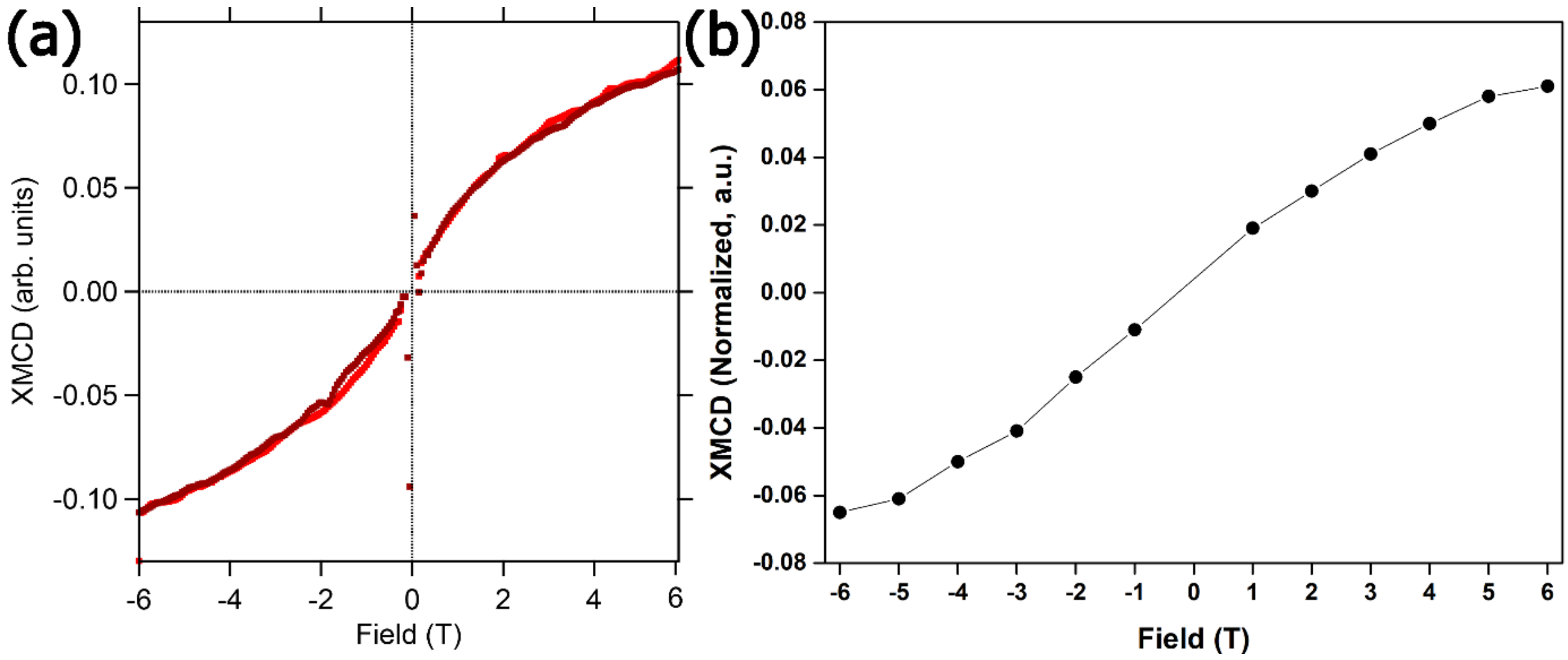

**Figure S11**. (a) XMCD hysteresis taken at the Cr $L_{2,3}$ edges at the 4 K temperature, (b) Field-dependent measurements of XMCD intensity for fields ranging from 6 T to − 6 T (T = 2 K).

## SUPPLEMENTARY MATERIALS FOR THE THEORY PART

### DFT Computational details

The spin-polarized density-functional theory (DFT) calculations were performed using the SIESTA 4.1.5 program[9,10]. The exchange-correlation potential was described within the Generalized Gradient Approximation (GGA) with the Perdew-Burke-Ernzerhof (PBE) parametrization. The core electrons were treated within the frozen core approximation, applying norm-conserving Troullier–Martins pseudopotentials. The wave functions of all Cr, I, and C valence states were described using the double-$\zeta$ polarized basis set. The *k*-point mesh was generated by the method of Monkhorst and Pack and with the cutoff of 15 Å for *k*-point sampling. The real-space grid used for the numeric integrations was set to correspond to the energy cutoff of 300 Ry. The calculations of the unit cells were performed with the global geometry optimization with convergence criteria corresponding to the maximum residual stress of 0.1 GPa for each component of the stress tensor, and the maximum residual force component of 0.05 eV/Å. For 1D and 2D nanostructures the lattice parameters in "non-periodic" directions were chosen as 40 Å to avoid interaction between periodic images.

Test calculations of $CrI_3$ single layer and its heterostructures with graphene within the simplified rotationally invariant DFT+U formalism[11] demonstrated only a marginal influence of the Coulomb repulsion on both the value and the composition of the fundamental bandgap, thereafter, $U_{eff}$ = 0 eV was employed only (**Figures S12** and **S13**). The possible spin-orbit effects were studied on optimized supercells and treated within on-site approximation for the spin-orbit matrix elements[12].

Preliminary geometry optimizations of two bulk $CrI_3$ polymorphs demonstrated a fair reliability of the chosen scheme in reproduction of the lattice parameters. Theoretical data $a$ = 6.79 Å, $b$ = 11.77 Å, $c$ = 6.74 Å, $\beta$ = 108.64 Å for monoclinic and $a$ = 6.82 Å, $c$ = 18.71 Å for hexagonal lattices are comparable to experimental ones $a$ = 6.87 Å, $b$ = 11.89 Å, $c$ = 6.98 Å, $\beta$ = 108.51 Å and $a$ = 6.87 Å, $c$ = 19.81 Å[7], respectively. Our additional calculations show that the obtained 6% underestimation of the interlayer distance can be reduced only to 4% using the vdW density functional of Dion-Rydberg at a larger computational cost, while a better refinement of interlayer distance has not been achieved by other authors even using the empirical Grimm D3 correction[13]. However, most of our theoretically considered nanostructures are single-layered, and superior description of $c$-parameter is overabundant here.

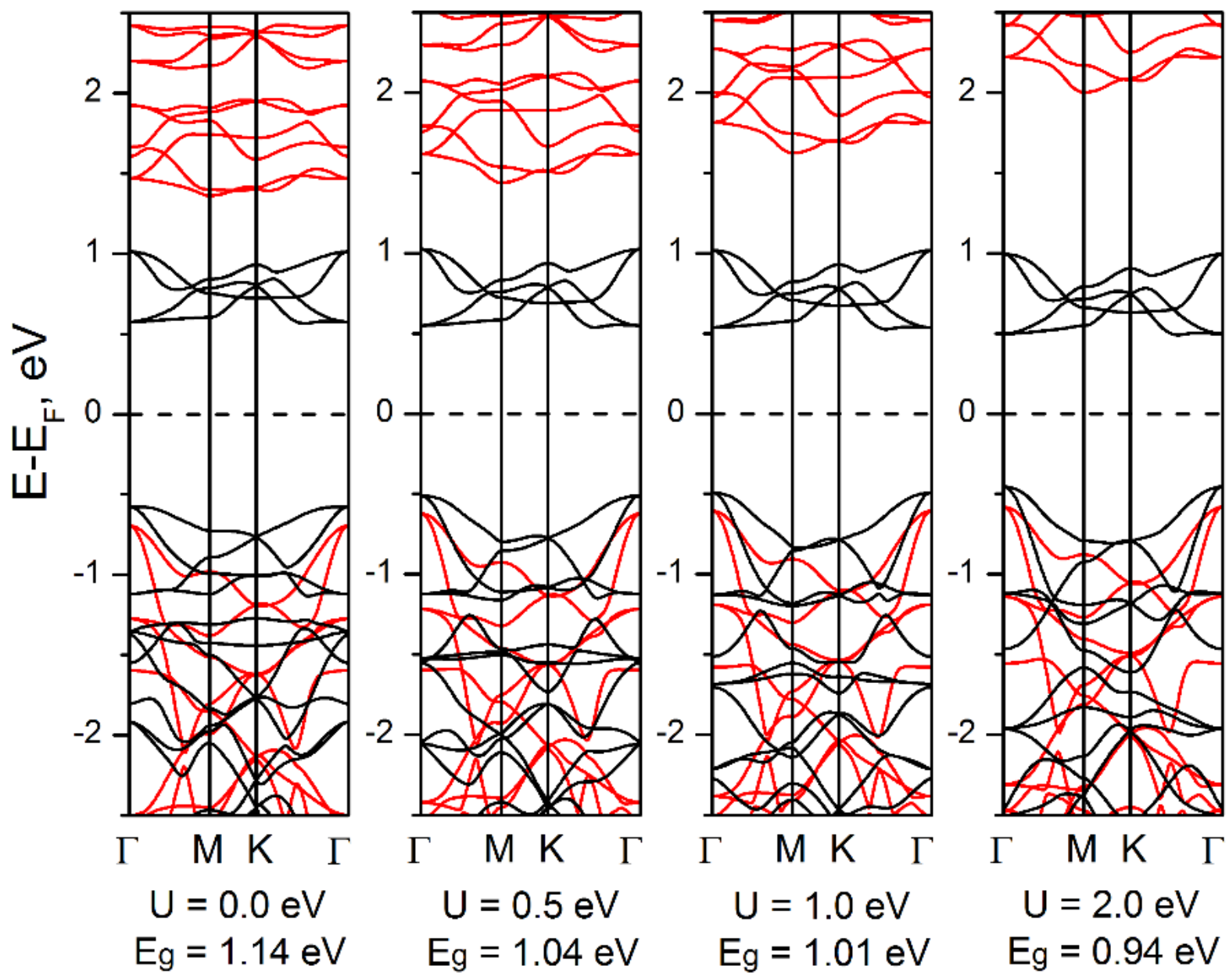

**Figure S12**. Band structure and the bandgap value ($E_g$) for freestanding single layer $CrI_3$ from DFT+U calculations as a function of $U_{eff}$ parameter. The dispersion curves for $\alpha$- and $\beta$-spin channels are painted in black and red, respectively.

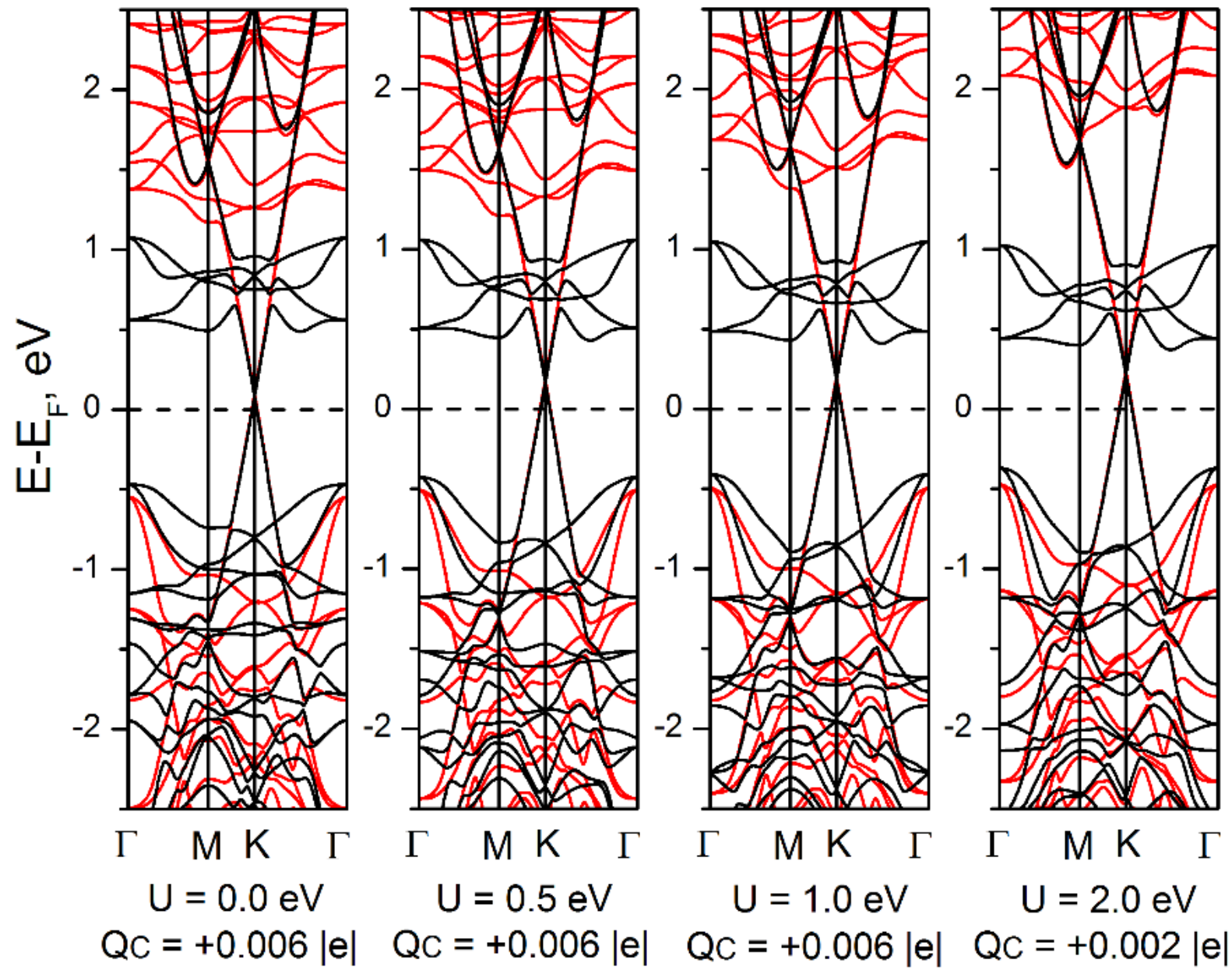

**Figure S13**. Band structure and the charge transfer ($Q_C$) from graphene to single layer $CrI_3$ within the 2D heterostructure (1×1)$CrI_3$-SL||(√7×√7)C-SL according to DFT+U calculations as a function of $U_{eff}$ parameter. The dispersion curves for α- and β-spin channels are painted in black and red, respectively.

**DFT Results**

***Structure and stability of free-standing $CrI_3$ nanostructures***

The relative stability of both curved and planar $CrI_3$ nanostructures has been analyzed using the density-functional theory (DFT) calculations. Geometry optimization of free-standing $CrI_3$ single layer shows evidence for preservation of its hexagonal symmetry with a slight in-plane contraction from $a$ = 6.82 Å to $a$ = 6.77 Å. Therefore, the models of the corresponding single-walled $CrI_3$ nanotubes can be folded using the same construction principles as for nanotubes of other hexagonal compounds[14]. We have studied nanotubes of two extreme chiralities: *armchair* ($n$,$n$) with $n$ = 2-8 and *zigzag* ($n$,0) with $n$ = 4-10 (**Figure S14**). Their stability and electronic properties have been compared to those for *armchair* ($n$,$n$) and *zigzag* ($n$,0) $CrI_3$ nanoribbons at $n$ = 1-3 with the different edges' constructions. The constructed *zigzag* nanoribbons had either glide- or mirror-symmetric edges, all having a coordination number of Chromium CN = 5 at the edges. *Armchair* nanoribbons also had either glide- or mirror-symmetric edges, yet, with either CN = 5 or CN = 4, respectively (**Figure S14**).

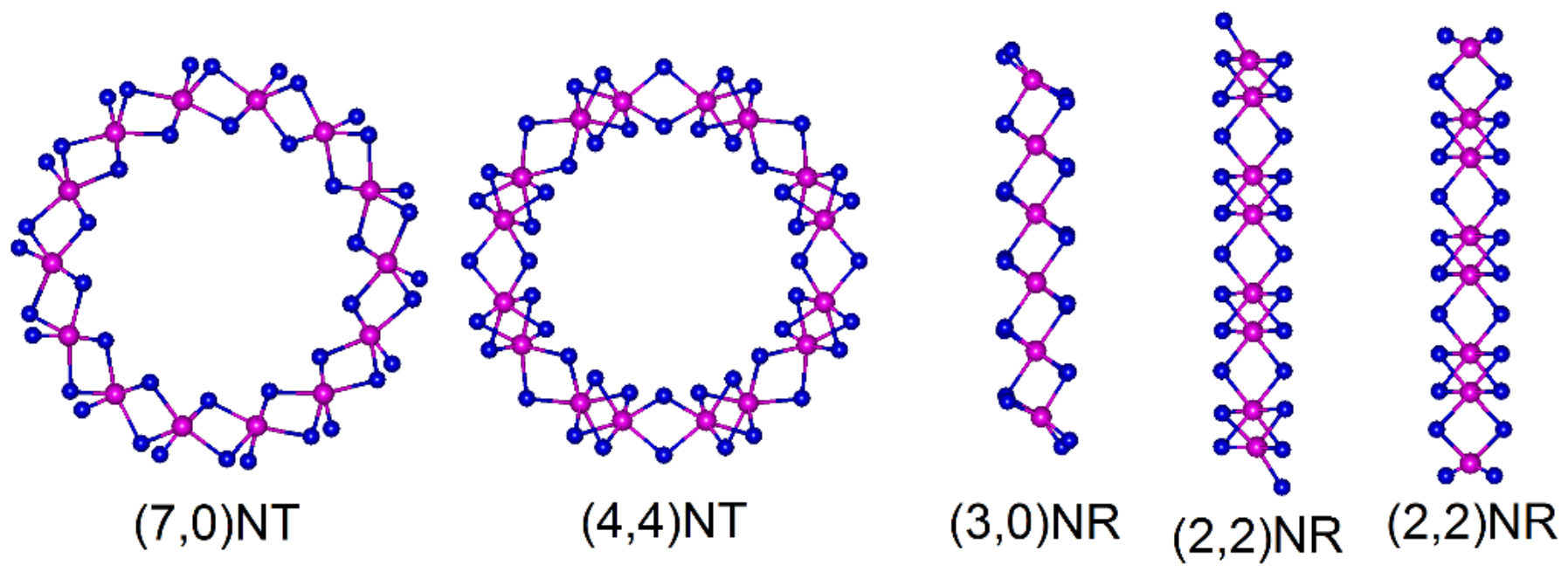

**Figure S14**. Ball-and-stick models of a few 1D single-walled $CrI_3$ nanostructures after geometry optimization using DFT method: *zigzag* (7,0) and *armchair* (4,4) nanotubes; *zigzag* (3,0) and two types of *armchair* (2,2) nanoribbons with either square-pyramidal or tetrahedral coordination number of the edge Cr atoms (CN = 5 or CN = 4). Views are along the main axes, Cr and I atoms are in violet and blue.

Initial cylindrical morphology is preserved after geometry reconstruction for all $CrI_3$ nanostructures irrespective of their chirality type in all range of chosen indices. The relative thermodynamic stability of nanotubes and flat nanoribbons as a function of both radius and chirality was traced using the energy ΔE calculated relative to the total energy of the single layer. For nanotubes this quantity should correspond to the pure strain energy, while that for nanoribbons should be contributed mostly from the energy of dangling bonds at the edges. The values of ΔE per $CrI_3$-unit are collected in **Figure S15**. Our data for nanotubes resemble closely the previous results[13,15]. These data demonstrate that the radius $R$ is the key factor governing the relative stability of cylindrical morphology of $CrI_3$. Irrespective of the chirality, the relative energy of nanotubes obeys the classical ~$1/R^2$ rule for the bending of a layer from the elasticity theory. The proportionality factor between ΔE and $1/R^2$ for all considered $CrI_3$ nanotubes is determined as 6.0 eV·Å$^2$/atom (24.06±0.19 eV·Å$^2$/unit), which is in-between the factors for nanotubes known for other layered compounds: it is above, than 3.0 eV·Å$^2$/atom for the topologically related $BiI_3$ nanotubes[16], and below, than ~16 eV·Å$^2$/atom for the commercially produced $MoS_2$ and $WS_2$ nanotubes[13,17]. Therefore, the calculations reveal, a mass fabrication of $CrI_3$ nanotubes is not impeded by the elastic strain of $CrI_3$ layers.

The lower bound of nanotubes' stability is predetermined by the energy of dangling bonds at the edges of corresponding nanoribbons. The values of ΔE per $CrI_3$-unit for every type of $CrI_3$ nanoribbons obey well the ~$1/R$ rule, where $R$ is the width of nanoribbon normalized by 2π, for every particular type of nanoribbon (**Figure S15**). This trend reflects an identical geometry reconstruction of edges as well as a negligible dependence of the energies of dangling bonds irrespective of the width of nanoribbons. Square-pyramidal coordination (CN = 5) of Cr atoms at the nanoribbons' edges seems to be more favorite, than tetrahedral one (CN = 4). However, the most important outcome from our DFT calculations consists in the cross-points between the ΔE functions for nanoribbons and nanotubes. Single walled $CrI_3$ nanotubes become thermodynamically more stable than nanoribbons, at radii larger, than 14 Å.

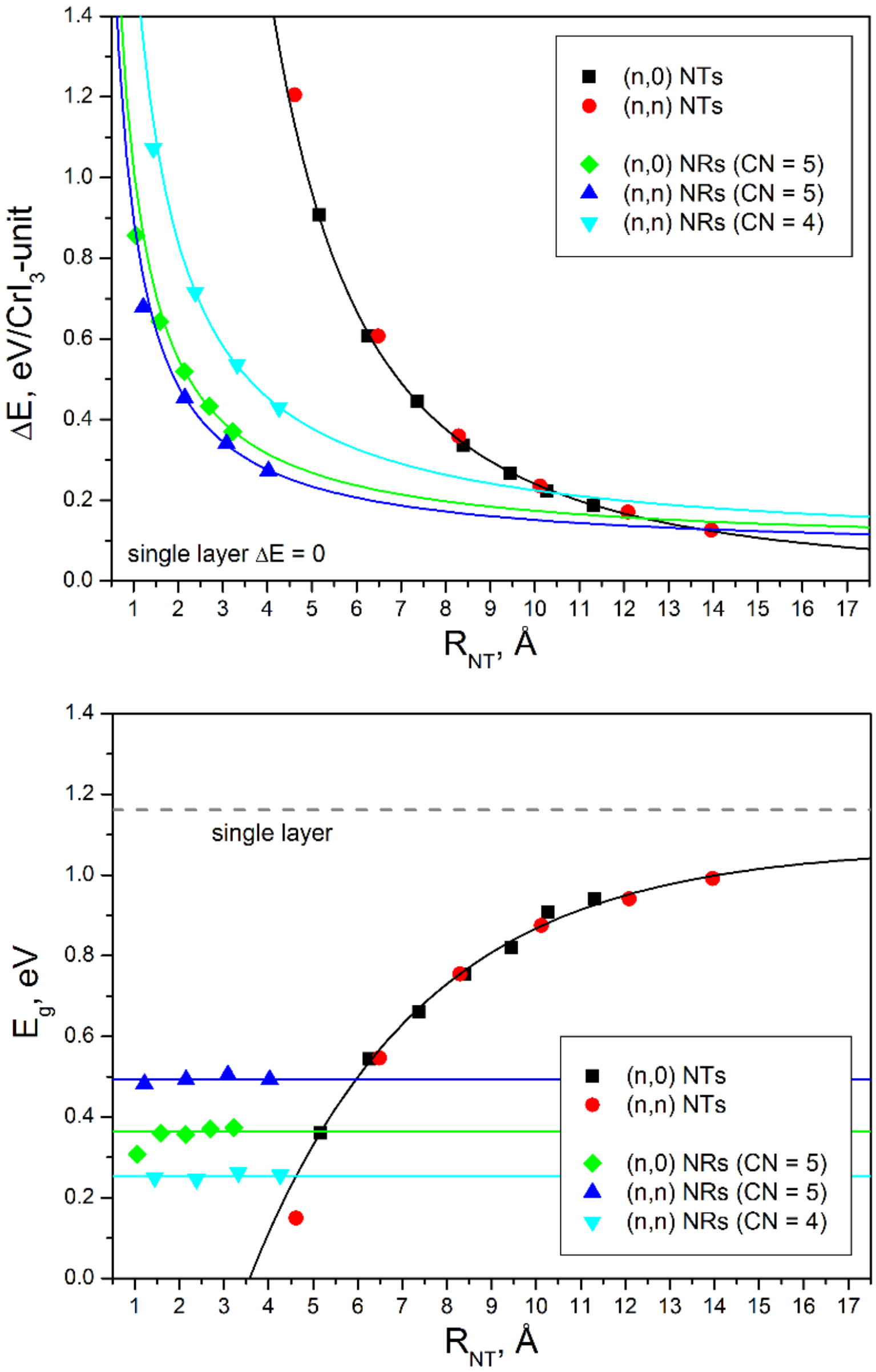

**Figure S15**. Relative energies ΔE and the bandgaps $E_g$ for different 1D $CrI_3$ nanostructures depending on their radii (for a nanoribbon the radius corresponds to the width normalized by 2π). The bandgap of nanotubes has a common origin with the fundamental bandgap of $CrI_3$ layer, while the bandgap of nanoribbons corresponds to the Highest occupied-lowest unoccupied (HOCO-LUCO) gap of the new states localized within the fundamental band gap of $CrI_3$ layer.

***Electronic properties of free-standing $CrI_3$ nanostructures***

DFT calculations allow trace a possible modulation of electronic properties of $CrI_3$ nanostructures as well as to explain the favorite edge construction for $CrI_3$ nanoribbons. Theoretical spin-resolved band structure and the density of electronic states (DOS) for the exemplary single-layer nanotubes and nanoribbons of $CrI_3$ are visualized in **Figure S16**. The distribution and relative position of the bands observed for different types of ferromagnetic nanostructures are qualitatively similar, exhibiting also a very close similarity with those of planar and infinite single layer. The top valence band and the conduction band consist of a mixture of I5*p*- and Cr3*d*-states, is responsible for the rise of polar covalent Cr-I bonding. In accordance to the crystal field theory every $Cr^{3+}$ ion in an octahedral field forms two groups of Cr3*d*-states. Cr3$d_{xz,yz,xy}$-levels for α-spin channel are fully populated by 3 electrons and their states can be found delocalized mostly at -0.5…-2.5 eV relative to the Fermi level. Unoccupied Cr3$d_{x2-y2,z2}$-levels for α-spin channel are separated by an energy gap from Cr3$d_{xz,yz,xy}$-levels and are localized within the conduction band at +0.5…+1 eV. All unoccupied Cr3*d*-levels for β-spin channel form a wide band at +1…+2.5 eV.

Table S2. Relative energies of ferromagnetic (FM) and antiferromagnetic (AFM) states calculated using DFT method for pristine $CrI_3$ nanotubes (NT), nanoribbons (NR) and single layer (SL) and for those in contact with carbon nanotubes and graphene. Positive values indicate a favour of FM state.

| Nanostructure | $E_{AFM}$ $E_{FM}$, meV/Cr-atom | Nanostructure | $E_{AFM}$ $E_{FM}$, meV/Cr-atom |
|---|---|---|---|
| (3,3)$CrI_3$NT | -3.0 | (3,3)$CrI_3$NT@(24,6)CNT | +13.0 |
| (5,0)$CrI_3$NT | +21.6 | (5,0)$CrI_3$NT@(22,11)CNT | +16.4 |
| (7,0)$CrI_3$NT | +17.4 | (7,0)$CrI_3$NT@(26,13)CNT | +14.8 |
| (2,2)$CrI_3$NR | +17.8 | (2,2)$CrI_3$NR@(24,6)CNT | +13.0 |
| (3,0)$CrI_3$NR | +17.1 | (3,0)$CrI_3$NR@(22,11)CNT | +16.7 |
| (1×1)$CrI_3$SL | +21.5 | (1×1)$CrI_3$SL‖(√7×√7)CSL | +2.7 |
| | | (√3×√3)$CrI_3$SL‖(5×5)CSL | +24.8 |

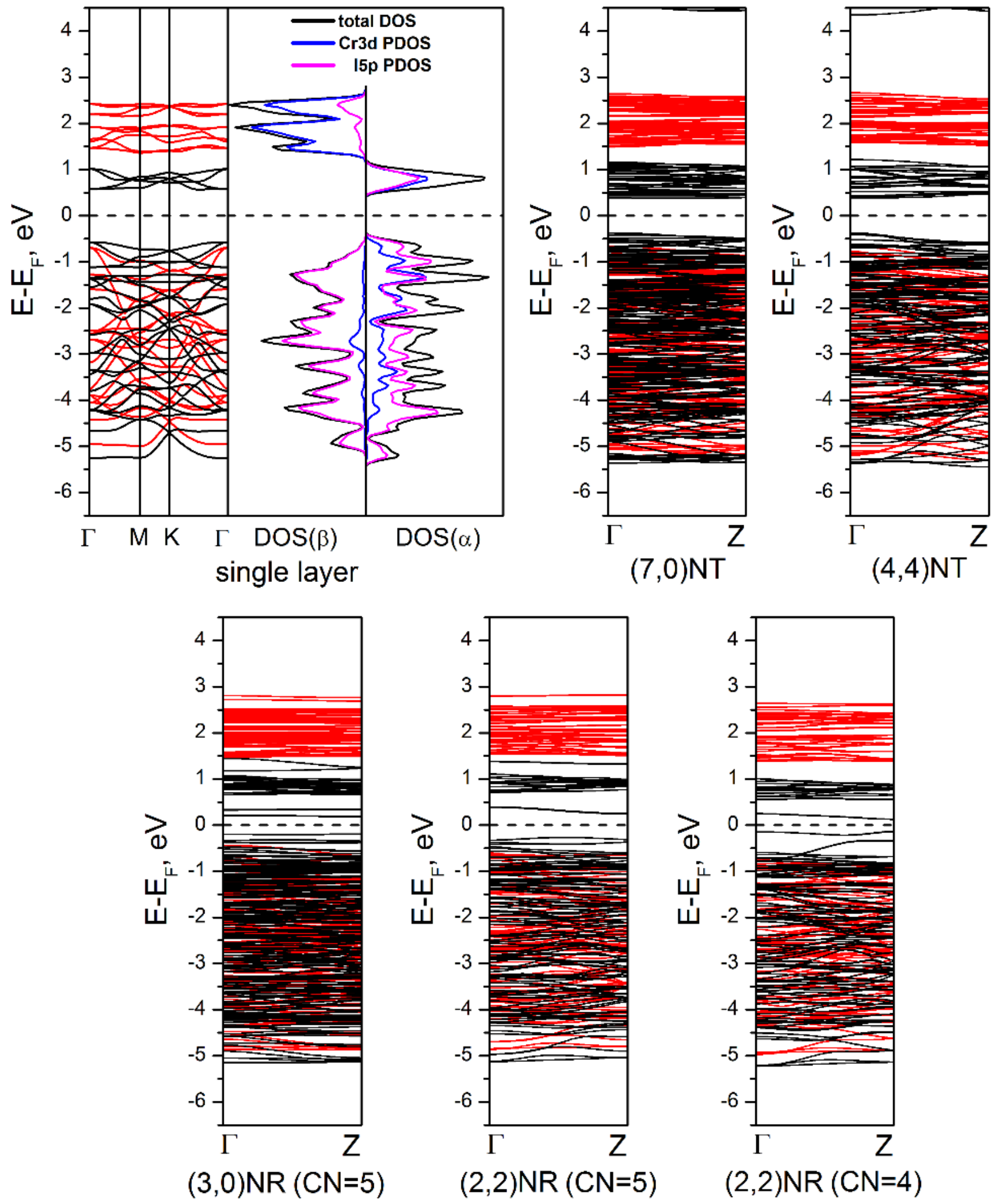

**Figure S16**. Band structure and density-of-states (DOS) for different freestanding $CrI_3$ nanostructures. For nanoribbons with different edges' construction the coordination number of the edge Cr atoms (CN) is given in brackets. The dispersion curves for α- and β-spin channels are painted in black and red, respectively.

All, but one, studied $CrI_3$ nanostuctures are ferromagnetic semiconductors (Table S2). For nanotubes, the value of the fundamental band gap $E_g$ decreases at decreasing diameter (**Figure S15**). Such a decreasing of the band gap of a seamless bent $CrI_3$ layer can be attributed to a facilitated overlap between the Cr3*d*- and I5*p*-orbitals at the internal side of a nanotube, at that no new states arise in the gap. At large radii, the values $E_g$ approach the value for single $CrI_3$ layer 1.14 eV. For nanoribbons, both the origin and the value of the fundamental bandgap remain nearly the same as those for single $CrI_3$ layer. Yet, their bandgap is perturbed by the rise of a combination of new Cr3*d*- and I5*p*-states, which are well localized and stem from the dangling bonds at nanoribbons' edges. The energy level difference between the new states as well as between these and the fundamental bands of $CrI_3$ are independent of the width of related nanoribbons. A low coordination of the edge Cr atoms causes another splitting of Cr3*d*-levels, than octahedral one, which also can be found in nice accordance to the schemes predicted using the classical crystal field theory. The Cr3*d*-level splitting within tetrahedral field is not gainful for population by 3 electrons of $Cr^{3+}$ ions: there are only two low-energy $d_{x2-y2,z2}$-levels. In contrast, the Cr3*d*-level splitting within square-pyramidal field is a reminiscence of octahedral field with three low-energy $d_{xz,yz,xy}$-levels, yet, with a higher energy of $d_{xy}$-level. The population of these three levels by 3 electrons of $Cr^{3+}$ ions is still energy gainful, yet, with a less gain, than for octahedral coordination. Therefore, in framework of crystal field theory, the new states localized in the fundamental band gap of a sliced $CrI_3$ layer are characterized mostly as $\alpha$-spin channels of the highest occupied crystal orbital $Cr3d_{xy}$ and the lowest unoccupied crystal orbital $Cr3d_{z2}$. As well, this theory explains the lower stability of $CrI_3$ nanoribbons with CN = 4 of the edge Cr atoms, than CN = 5.

***Structure and stability of $CrI_3$–C nano-heterostructures***

Further, DFT calculations have been employed to get an insight into the interaction degree and the possible electronic properties of heterostructures assembled of single-walled nanostructures of $CrI_3$ and carbon. Since the size of the fabricated 1D heterostructures between carbon nanotubes and $CrI_3$ exceeds overwhelmingly the computational possibilities here, their model set was limited to “extreme” variants of curvature existing in either the smallest possible coaxial $CrI_3$ and carbon nanotubes ($CrI_3$@C) or the so-called vertical heterostructures of planar $CrI_3$ single layer and graphene ($CrI_3$||C). Theoretical ratio between the in-plane parameters of the honeycomb lattices of graphite and bulk $CrI_3$ has been determined in our study as $a_C/a_{CrI3}$ = 2.48/6.82 = 0.364, which is close to experimental ratio $a_C/a_{CrI3}$ = 2.46/6.87 = 0.358. Therefore, the lowest mismatch between these lattices at an appropriate size of the model 2D $CrI_3$||C heterostructure can be achieved within the supercell composed of (1×1) $CrI_3$ layer cell and (√7×√7) graphene cell (in total 22 atoms) (**Figure S17**). Noteworthy, earlier DFT studies of $CrI_3$||C employing a plane-wave basis set have attributed the lowest mismatch for another supercell as of (√3×√3) $CrI_3$ layer cell and (5×5) graphene cell (in total 74 atoms)[18,19]. We have also included the latter supercell in consideration, though, it is characterized as a more stressed one in our LCAO DFT study.

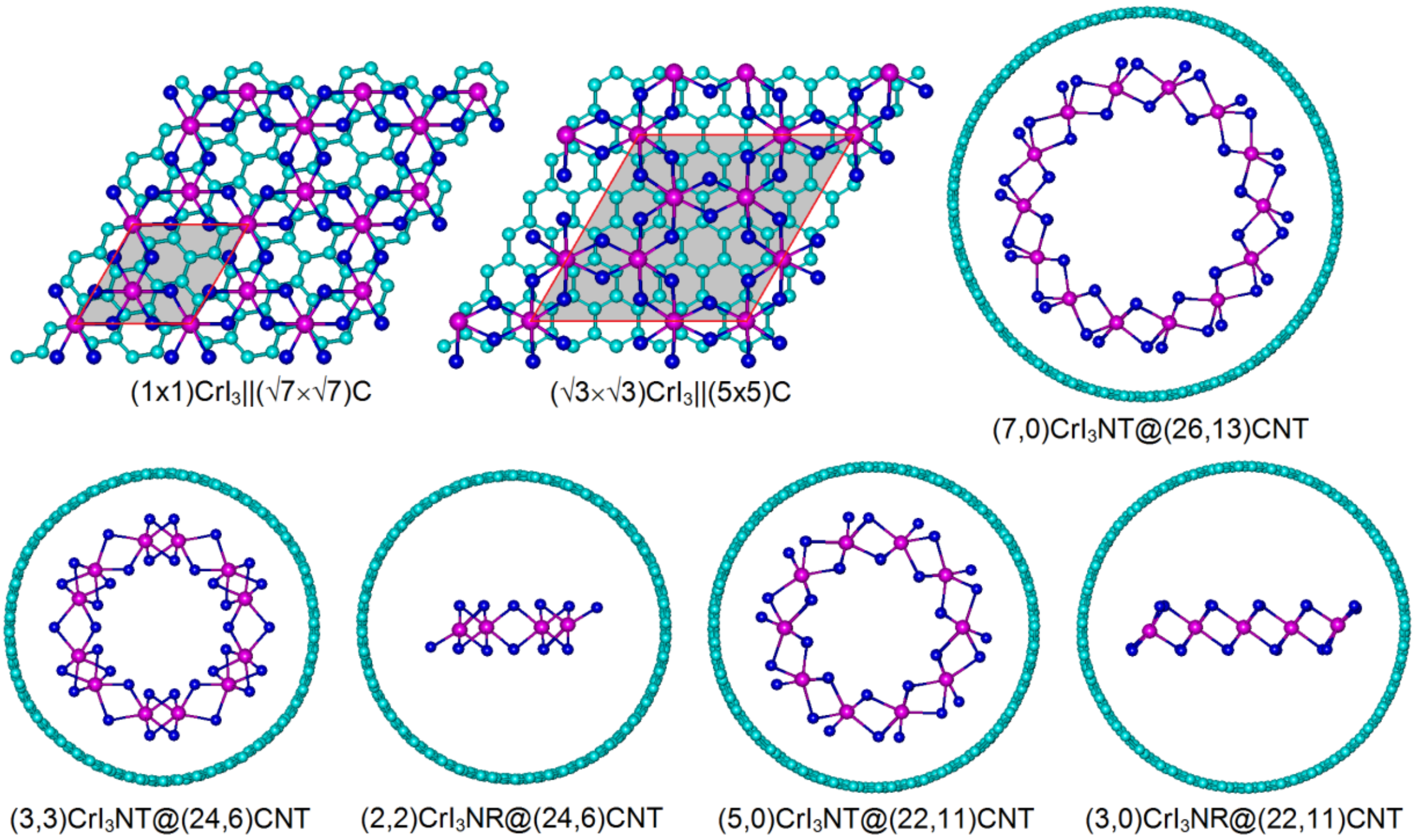

**Figure S17**. Ball-and-stick DFT-optimized models of the heterostructures composed of various single walled $CrI_3$ and single walled carbon nanostructures – planar layers, nanoribbons or cylindrical nanotubes (Cr, I and C atoms are in violet, blue and cyan, respectively). 2D heterostructures are viewed on their basal planes, while 1D heterostructures are viewed along their main axes. The unit cells of 2D heterostructures are painted.

The lattice parameters of 2D $CrI_3$||C ($CrI_3$/graphene) heterostructures after geometry optimization resemble that of the carbon part with only 0.3% expansion, which reflects a supremacy of the Young´s modulus of graphene under periodic boundary conditions. The formation energy calculated relative to the free-standing and not-stressed layers is found exothermic and equal to -17.5 meV/Å$^2$ for the most stable interface (1×1)$CrI_3$||(√7×√7)C. The distance between Cr and C planes is optimized to ~5.0 Å.

The choice of DFT models of 1D $CrI_3$@C nanotubular heterostructures is complicated by balancing between the lattices' match along both axial and radial directions and the atomistic size. Among rich variety of chiralities the (24,6), (22,11) and (26,13) carbon nanotubes have been selected, which can beget $CrI_3$@C supercells of 216, 388 and 476 atoms by the (3,3), (5,0) and (7,0) $CrI_3$ nanotubes, respectively (**Figure S17**). Such selection provides a tolerable lattice mismatch of ~3% in axial direction for all optimized $CrI_3$@C nanostructures: $CrI_3$ part is "in-plane compressed" under periodic boundary conditions similarly to $CrI_3$||C heterostructures, while interplane Cr-C distances are equal to 4.7, 5.1 and 5.2 Å in (3,3)$CrI_3$@(24,6)C, (5,0)$CrI_3$@(22,11)C and (7,0)$CrI_3$@(26,13)C. As well, the contact of curved $CrI_3$ layers to both metal and semiconducting carbon nanotubes can be analyzed.

The calculations in general approve a preservation of the FM state of $CrI_3$ within all heterostructures irrespective of both the morphology of $CrI_3$ and the electronic properties of the carbon part: in most cases, a weakening of ferromagnetic coupling is obtained since the energy of FM state for a $CrI_3$ component decreases on 0.4-5.2 meV/Cr-atom after contact with a carbon (Table S3). Therefore, electronic properties of all $CrI_3$-C heterostructures will be discussed *vide infra* for the FM case only. Here, two attention-grabbing systems can be noticed. First, the ground magnetic state for extremely curved (3,3) $CrI_3$ was found as AFM. Yet, it becomes FM within a CNT. It can be explained by a large radial expansion of the $CrI_3$ layer as free-standing (3,3)$CrI_3$ NT, while a rigid wall of CNT is capable of preventing such expansion and does stabilize interatomic distances within $CrI_3$ closer to those within flat $CrI_3$ layer: Cr-Cr distances in (3,3)$CrI_3$NT, in (3,3)$CrI_3$NT@(24,6)CNT and in $CrI_3$SL are equal 4.30 Å, 4.15 Å, and 3.91 Å, respectively. Second, the FM state of the flat $CrI_3$ layer gets drastically weaker within (1×1)$CrI_3$||(√7×√7)C heterostructure. Obviously, it is related to unique stacking between the $CrI_3$ layer and graphene here, where Cr-sublattice separates into two groups of Cr-atoms lodged strictly either on C-atoms or on C-hexagons. Such a registry between $CrI_3$ and graphene suppresses the degeneracy of $Cr^{3+}$ ions in the AB alternation. However, this phenomenon disappears at other stacking types, where a richer diversity of Cr-atoms in both planar and cylindrical heterostructures does exist.

Table S3. Charge transfer from a carbon to a $CrI_3$ nanostructure ($Q_C$) was calculated using the DFT method without and with spin-orbit coupling (SOC) within various $CrI_3$@CNT and $CrI_3$||C heterostructures. N – numbers of $CrI_3$ units or C atoms within the supercells used.

| Nanostructure | N($CrI_3$) | N(C) | $Q_C$, \|e\| (no SOC) | $Q_C$, \|e\| (SOC) |
|---|---|---|---|---|
| | | stoichiometric compositions | | |
| (3,3)$CrI_3$NT@(24,6)CNT | 12 | 168 | +0.186 | +0.207 |
| (5,0)$CrI_3$NT@(22,11)CNT | 20 | 308 | +0.531 | +0.414 |
| (7,0)$CrI_3$NT@(26,13)CNT | 28 | 364 | +0.322 | +0.186 |
| (2,2)$CrI_3$NR@(24,6)CNT | 4 | 168 | +0.030 | +0.001 |
| (3,0)$CrI_3$NR@(22,11)CNT | 10 | 308 | +0.287 | +0.052 |
| (1×1)$CrI_3$SL\|\|(√7×√7)CSL | 2 | 14 | +0.006 | +0.006 |
| (√3×√3)$CrI_3$SL\|\|(5×5)CSL | 6 | 50 | +0.055 | +0.054 |
| | | non-stoichiometric compositions | | |
| (5,0)$CrI_{3-\delta}$NT@(22,11)CNT | $Cr_{20}I_{59}$ | 308 | +0.600 | bad conv. |
| (1×1)$CrI_{3-\delta}$SL‖(√7×√7)CSL | $Cr_2I_5$ | 14 | +0.003 | +0.004 |
| (√3×√3)$CrI_{3-\delta}$SL‖(5×5)CSL | $Cr_6I_{17}$ | 50 | +0.173 | +0.146 |

***Electronic properties of $CrI_3$–C one-dimensional structures***

The band structure of (1×1)$CrI_3$‖(√7×√7)C reveals a spin-dependent hybridization the graphene Dirac cone with the spin-polarized conduction band of $CrI_3$, in agreement with results reported in the literature[20]. In addition, our calculations also predict charge transfer from the CNT to the $CrI_3$ (**Figure S18**, **Table S3**). The Fermi level appears in the extreme vicinity of the Dirac point formed by the conduction and valence bands of C2*p*-states of graphene, while the Dirac cones are clearly visible within the fundamental band gap of $CrI_3$. The same features of the band structure are acquired by (√3×√3) $CrI_3$‖(5×5)C heterostructure, yet, with the Dirac point of graphene shifted closer to the bottom conduction band of $CrI_3$. Such a reorganization of the levels may be explained by a higher lattices' mismatch, hence, a higher internal stress of this particular heterostructure constrained under periodic boundary conditions. In general, our results on the band structure of 2D heterostructures using LCAO DFT are in line with the results from aforementioned DFT studies using a plane-wave basis set[18,19,21], disagreeing only in the exact alignment of the Dirac point from graphene with the conduction band of stoichiometric $CrI_3$: here, the position of the Dirac point is predicted to align below the unoccupied Cr3*d*-band, not in the middle. Like in a previous plane-wave DFT study[21], $CrI_3$‖C heterostructure may be qualified as a typical metal-semiconductor contacted system of the van der Waals' bounded layers. Charge transfer is associated to the work-function mismatch of $CrI_3$ and graphene/nanotubes (Table S3).

The electronic structure of (3,3)$CrI_3$NT@(24,6)CNT unveils its metallic properties: the Dirac cones of parent (24,6)CNT are hosted within the fundamental band gap of (3,3)$CrI_3$NT nanotube with a small ~0.03 eV gap opening (**Figure S18**). The band structure of heterostructure represents mostly a superposition of the band structures of both components. The hybridization between C2*p*- and Cr3*d*-states is rather weak and the former Dirac point does not align with the conduction band of $CrI_3$ even at this high curvature of $CrI_3$ layer. The Fermi level is shifted down to the valence band of C2*p*-states. We find charge transfer from carbon to $CrI_3$ nanotube can be registered (~0.2 e per supercell, Table S3). The band structures of the parent components can also be easily distinguished in the band structures of both (5,0)$CrI_3$NT@(22,11)CNT and (7,0)$CrI_3$NT@(26,13)CNT heterostructures, where all the individual components are semiconductors. Electronic properties of these 1D heterostructures can be described as for a semi-metal or for a semiconducting heterojunction of type II (staggered gap) with the zero bandgap: the bottom of conduction band is given by $CrI_3$ conduction band, while the top of valence band is given by a carbon semiconductor. Charge transfer is also found in these heterostructures, too.

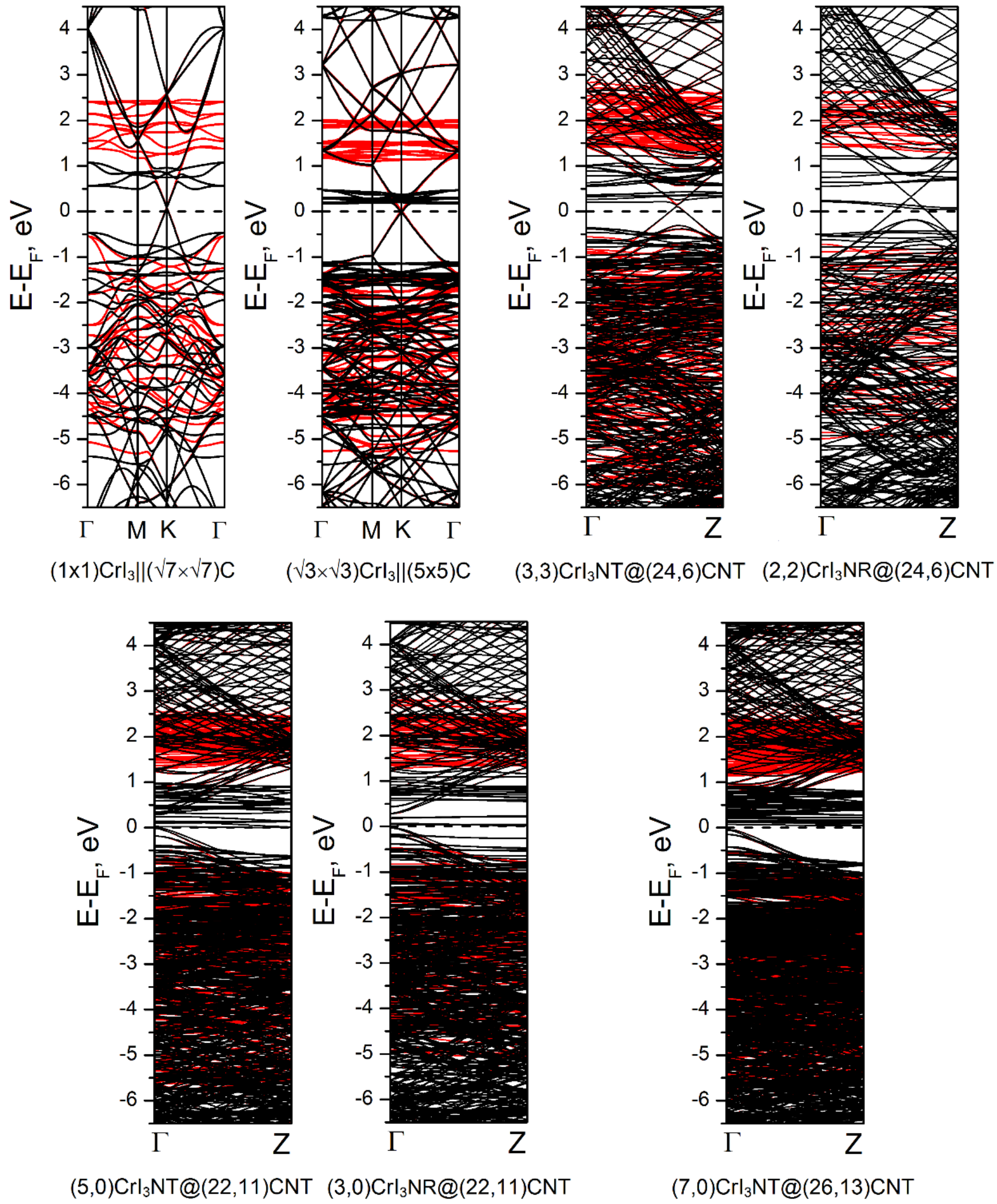

**Figure S18**. Band structure of different nano-heterostructures composed of single-walled carbon and single-walled $CrI_3$ parts (ball-and-stick models are shown in **Figure S17**). The dispersion curves for α- and β-spin channels are painted in black and red, respectively. DFT calculations.

In addition, apart from $CrI_3$NT@CNT, the smallest 1D heterostructures as CNTs filled by single layer $CrI_3$ nanoribbons have been also designed, preserving the minimal Cr-C distance at ~5 Å as in $CrI_3$NT@CNT and $CrI_3$||C heterostructures (**Figure S18**). Irrespective of the presence of the edges in the flat $CrI_3$ layer, the hybridization between C2*p*- and Cr3*d*-states remains weak. The (2,2)$CrI_3$NR@(24,6)CNT heterostructure is metallic due to the parent (24,6)CNT (**Figure S17**). The Fermi level is shifted down to the C2*p*-band, while both Cr3*d*- states from dangling bonds and C2*p*-cone are well localized within the fundamental band gap of $CrI_3$. The (3,0)$CrI_3$NR@(22,11)CNT heterostructure is a semiconductor with the bandgap, which arises from the bandgap of the parent (22,11) carbon nanotube, hosting the strongly localized Cr3*d*-states from the edge atoms of $CrI_3$ nanoribbon. Some charge transfer is also obtained in these heterostructures irrespective of the origin of parent components (Table S3).

Due to both stress at the interior of $CrI_3$ nanotubes and a high fugacity of iodine, the presence of I vacancies cannot be excluded. In principle, the rise of I vacancies could lead to a modification of the electronic structure similar to those for $CrI_3$ nanostripes, i.e. appearance of the states localized within the fundamental bandgap of $CrI_3$. Our additional DFT calculations confirm the rise of the mixed Cr3*d*-I5*p* levels, which are indeed localized in the bandgap of $CrI_3$ and near the top of the valence C2*p*-band of a semiconducting carbon or at the Dirac point of a metallic carbon (**Figure S18**). The nominal charge transfer within an I-deficient heterostructure may be larger than that within a stoichiometric heterostructure, yet it remains small and is caused by redistribution of electronic density at the interface (Table S3). A more detailed analysis shows evidence that excessive electron density is delocalized mostly between the I atoms.

The aforementioned peculiarities in the electronic structure of $CrI_3$@C and $CrI_3$||C heterostructures with both stoichiometric and I-deficient compositions are mostly preserved in DFT calculations accounting for spin-orbit coupling (**Figure S20**). In some cases, the occupied I5*p*-states can be shifted toward the Fermi level, yet it does not qualitatively change the conductivity type and the degree of charge transfer within these nanostructures.

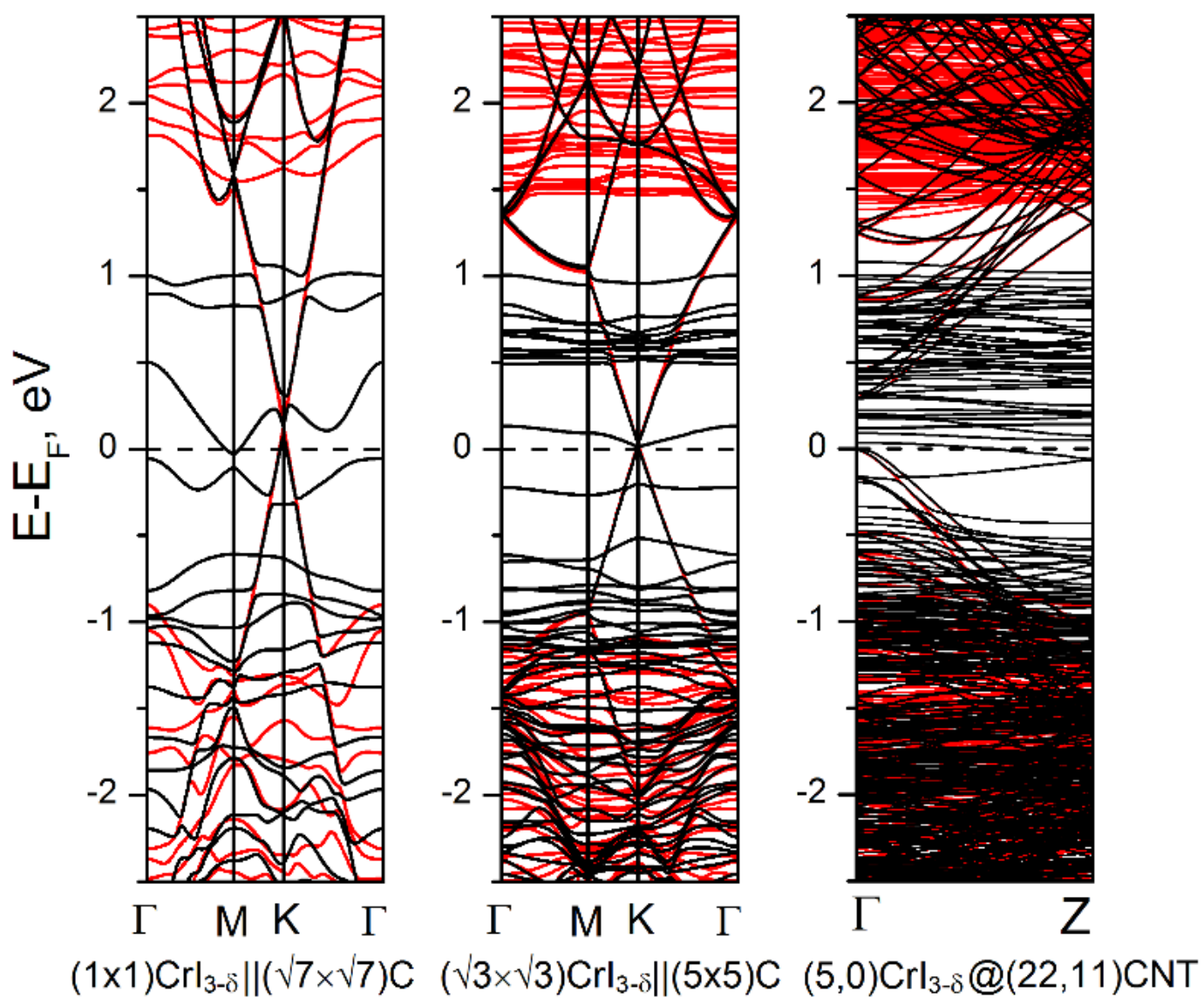

**Figure S19**. Band structures for several 2D- and 1D-nanoheterostructures of carbon and $CrI_3$ with single I vacancy. The exact δ can be calculated using data in Table S3. The dispersion curves for α- and β-spin channels are painted in black and red, respectively.

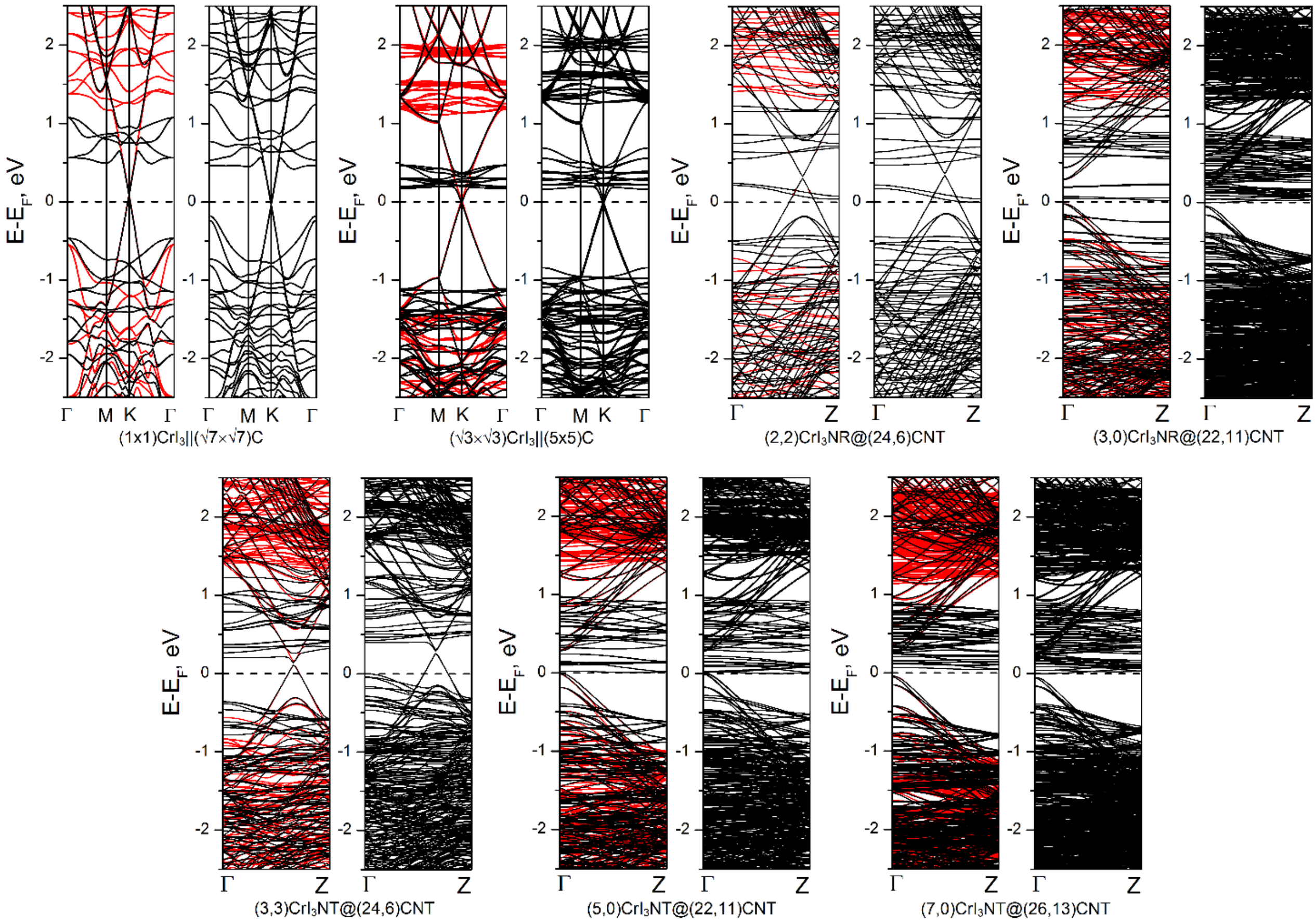

**Figure S20**. Comparison between the band structures of $CrI_3$||C and $CrI_3$@CNT heterostructures calculated using the DFT method without and with the account of spin-orbit coupling for all elements (left and right panels at any particular system, respectively). The dispersion curves for α- and β-spin channels on the left panels are painted in black and red.